\documentclass[12pt]{article}
\usepackage{graphicx, here, amsfonts, epsfig, hhline}
\def\Journal#1#2#3#4{{#1} {\bf #2}, #3 (#4)}

\def\YAD{\em Yad. Fiz.}
\def\NCA{\em Nuovo Cimento}

\def\NPB{{\em Nucl. Phys.} B}
\def\PLB{{\em Phys. Lett.}  B}

\def\PRD{{\em Phys. Rev.} D}
\def\ZPC{{\em Z. Phys.} C}
\def\EPG{{\em Eur.Phys.J.} C}

\def\AOP{{\em Ann. of Phys. (NY)}}

\def\be{\begin{equation}}
\def\ee{\end{equation}}
\def\bea{\begin{eqnarray}}
\def\eea{\end{eqnarray}}
\def \Pom {{\hspace{ -0.05em}I\hspace{-0.25em}P}}
\def \Reg {{\hspace{ -0.05em}I\hspace{-0.25em}R}}

\begin{document}
\begin{center}
{\Large \bf Photoproduction of vector mesons in the Soft Dipole
Pomeron model}

\vskip 1.cm {E. MARTYNOV$^{a,b,}$\footnote{\it E-mail:
E.Martynov@guest.ulg.ac.be},
 E. PREDAZZI$^{c,}$\footnote{\it E-mail: predazzi@to.infn.it}
 and A. PROKUDIN$^{c,d,}$\footnote{\it E-mail: prokudin@to.infn.it}}
\vskip 0.5cm
{\small\it
(a) Bogolyubov Institute for Theoretical Physics,\\ National Academy of
Sciences of Ukraine, \\ 03143 Kiev-143, Metrologicheskaja 14b, UKRAINE}
\vskip 0.2cm
{\small\it
\vskip 0.2cm
(b)  Institut de physique Bat B5-a \\
Universit\'e de Li\`ege \\
Sart Tilman B-4000 Li\`ege, \\
  BELGIQUE
}
{\small\it
\vskip 0.2cm
(c)  Dipartimento di Fisica Teorica,\\
Universit\`a Degli Studi Di Torino, \\
Via Pietro Giuria 1,
10125 Torino, \\
ITALY\\
and\\
Sezione INFN di Torino,\\
 ITALY\\}
\vskip 0.2cm
{\small\it
(d) Institute For High Energy Physics,\\
142281 Protvino,  RUSSIA}
\vskip 0.5cm
\parbox[t]{12.cm}{\footnotesize Exclusive photoproduction of all
vector mesons by real and
virtual photons is considered in the Soft Dipole Pomeron model.
It is emphasized that being the Pomeron in this model a double Regge
pole with intercept equal to one, we are led to rising cross-sections but
the unitarity bounds are not violated. It is shown that
all available data for $\rho, \omega, \varphi, J/\psi $ and
$\Upsilon $ in the region of energies 1.7 $\leq W \leq $ 250 GeV and
photon virtualities 0 $\leq Q^2 \leq $ 35 GeV$^2$ , including
the differential cross-sections in the region of transfer momenta
 0 $\leq |t| \leq$ 1.6 GeV$^2$, are well described
by the model. }
\end{center}

\section{Introduction}

A new precise measurement of $J/\psi$ exclusive photoproduction
by ZEUS \cite{NEWZEUS} opens a new window in our understanding
 of the process and allows us to give more
accurate predictions for future experiments.

The key issue of the dataset \cite{NEWZEUS} is the diffractive
cone shrinkage observed in  $J/\psi$ photoproduction which
leads us to consider it a soft rather than pure QCD
process so that we can apply the Soft
Dipole Pomeron exchange \cite{owrmodel} model.

We are improving the model while not changing its main
properties such as the universality for all vector mesons and its
applicability in a wide energy region. The structure of the
amplitude singularities in the $j$-plane remains also intact but
we use a nonlinear Pomeron trajectory in order to describe
correctly the behaviour of the differential distributions. The
use of non-linear trajectories improves, in fact, the
analyticity properties of the scattering amplitude. The
secondary Reggeons however,
 are for simplicity, taken directly with their
trajectories as determined from the pure hadronic case. The $J/\psi$
elastic cross section is described as due to the soft Pomeron
exchange but without unitarity violation.

We utilize the following picture of the interaction: a photon
fluctuates into a quark-antiquark pair and as the lifetime of
such a fluctuation is quite long (by the uncertainty principle
it grows with the beam energy $\nu$ as $2\nu/(Q^2+M_V^2)$
\cite{ref:ioffe}), the proton interacts via Pomeron or secondary
Reggeon exchange with this quark-antiquark pair. After the
interaction this pair forms a vector meson \cite{ref:Brodsky}.
The hint is that such an interaction must be very close to
that among hadrons and, following the principle of Regge pole
theory, that the Pomeron is universal in all hadron-hadron
interactions and
in all other processes, including DIS, provided we have
an appropriate kinematical region for the Regge approach to hold
(vacuum quantum numbers exchange is possible).
Thus, if Pomeron exchange
is possible,
then it has the same properties (the form of singularity,
position of such a singularity in the $J$-plane, trajectory
etc.) as in hadron-hadron interaction. This is true at least for
on shell particles. A real photon ($Q^2=0$) is considered
it as a hadron (according to the data). For
$Q^2\neq 0$ we assume that no new singularity appears
\cite{ref:Evolution}. More precisely, even if we assume a new
singularity at $Q^2\neq 0$, its contribution must be equal to
zero for $Q^2= 0$. Indeed the analysis of the data
\cite{ref:Desgrolard} shows that there is no need for such a new
contribution.

\begin{figure}[h]
\begin{center}
\includegraphics*[scale=0.4]{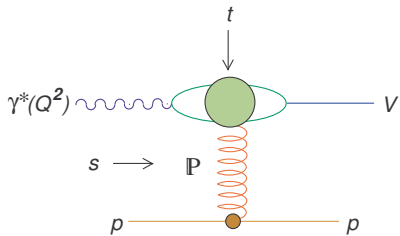}
\caption{Photoproduction of a vector meson.}
\label{Figure 1}
\end{center}
\end{figure}

The basic diagram is depicted in Figure \ref{Figure 1}; $s$ and $t$ are
the usual Mandelstam variables, $Q^2=-q^2$ is the virtuality of the
photon.

It is well known that high-energy representation of the
scattering amplitude may be expressed as a sum over the
the appropriate Regge poles in the complex $j$ plane
\cite{ref:PredazziBarone}

\be
A(s,t)_{s\to \infty}\approx \sum\limits_{i}
\eta_{i}(t)\beta_{i}(t)(\cos\theta_{t})^{\alpha_{i}(t)},
\ee
where $\eta_{i}(t)$ is the signature factor and $\theta_t$ is the scattering angle
in the $t$ channel.

In the case of vector meson photoproduction we utilize the
variable $z \sim \cos \theta_t$ \be z =
\frac{2(W^2-M_p^2)+t+Q^2-M_V^2}{\sqrt{(t+Q^2-M_V^2)^2+4M_V^2Q^2}}
\ee where  $W^2=(p+q)^2\equiv s$, $M_V$ is the vector meson
mass, $M_p$ is the proton mass.

 Assuming vector meson dominance \cite{ref:vdm}, the relation between
the forward cross sections of $\gamma p\rightarrow V p$ and $V
p\rightarrow V p$ is given by \be
\frac{d\sigma}{dt}(t=0)_{\gamma p\rightarrow V p}=
\frac{4\pi\alpha}{f_V^2}\frac{d\sigma}{dt}(t=0)_{V p\rightarrow
V p} \ee where the strength of the vector meson coupling
$\frac{4\pi}{f_V^2}$ may be found from $e^+e^-$ decay width of
vector meson $V$ \be \Gamma_{V\rightarrow e^+e^-}=
\frac{\alpha^2}{3}\frac{4\pi}{f_V^2} m_V \ee When  $V=\rho_0,
\omega, \varphi, J/\Psi$ the relations of these couplings may be
obtained assuming SU(4) flavour symmetry. No attempt is made to
extend flavour symmetry to SU(5) so as to incorporate also the
$\Upsilon$ coupling. The symmetry is too badly broken for this
to make sense. Using the quark content of the mesons, we have
\be
\begin{array}{lll}
\displaystyle &<Q^2_j>_\rho &= \Big|
\frac{1}{\sqrt{2}}(\frac{2}{3}+\frac{1}{3})\Big|^2=\frac{1}{2}\;
,\\ \nonumber \displaystyle &<Q^2_j>_\omega &=
\Big|\frac{1}{\sqrt{2}}(\frac{2}{3}-\frac{1}{3})\Big|^2=\frac{1}{18}\;
,\\ \nonumber \displaystyle &<Q^2_j>_\phi &=
\Big|\frac{1}{3}\Big|^2=\frac{1}{9}\; ,\\ \nonumber
\displaystyle &<Q^2_j>_{J/\psi} &=
\Big|\frac{2}{3}\Big|^2=\frac{4}{9}\; . \nonumber
\end{array}
\ee

Using the property $\Gamma_{V\rightarrow e^+e^-}/<Q^2_j>\simeq const$
we can obtain the following
approximate relations

\be
{m_\rho/f_\rho^2}:{m_{\omega}/f_{\omega}^2}:{m_{\varphi}/f_{\varphi}^2}:
{m_{J/\psi}/f_{J/\psi}^2} =
{9}:{1}:{2}:{8}\;
\ee
which are in fairly good agreement with experimental measurements
of decay widths \cite{ref:groom}.

We take into account these relations by introducing coefficients
$N_V$ (following to \cite{ref:Nemchik} ) and writing the
amplitude as $A_{\gamma p\rightarrow V p}=N_C N_V A_{V
p\rightarrow V p}$, where \be N_C = 3\: ; N_\rho =
\frac{1}{\sqrt{2}}\: ; N_\omega = \frac{1}{3\sqrt{2}}\: ; N_\phi
= \frac{1}{3}\: ; N_{J/\psi} = \frac{2}{3}\: . \ee

The
amplitude of the process $V p\rightarrow V p$
may be written in the
following form \be
A(z,t;M_V^2,\tilde Q^2) = \Pom (z,t;M_V^2, \tilde Q^2) + f(z,t;
M_V^2,\tilde Q^2) + ...\;, \ee where,
$\tilde Q^{2} = Q^{2}+M_V^2$.  

$\Pom (z,t;M_V^2, \tilde
Q^2)$ is the Pomeron contribution for which we use the so called
dipole Pomeron which gives a very good description of all
hadron-hadron total cross
sections \cite{ref:dipole_hadron},\cite{ref:COMPETE}.
Specifically, $\Pom$ is given by  \cite{ref:JMP} \be \Pom
(z,t;M_V^2, \tilde Q^2) = ig_{0}(t;M_V^2, \tilde Q^2)
(-iz)^{\alpha_{\Pom}(t)-1} + ig_{1}(t;M_V^2, \tilde Q^2)ln(-iz)
(-iz)^{\alpha_{\Pom}(t)-1}\; , \label{eq:pomeron} \ee where the
first term is a single $j$-pole contribution and the second
(with an additional $ln(-iz)$ factor) is the contribution of the
double $j$-pole.

A similar
expression applies to the contribution of the $f$-Reggeon 
\be
f(z,t;M_V^2, \tilde Q^2) = ig_{f}(t;M_V^2,\tilde Q^2)
(-iz)^{\alpha_{f}(t)-1}. 
\label{eq:reggeon} 
\ee

It is important to stress that in this model the intercept of
the Pomeron trajectory is equal to 1 \be \alpha_{\Pom}(0) = 1.
\ee Thus the model does not violate the Froissart-Martin bound
~\cite{ref:MartinF}.
 
For $\rho$ and $\varphi$ meson photoproduction we write the
scattering amplitude as the sum of a Pomeron and $f$
contribution. According to the Okubo-Zweig rule, the $f$ meson
contribution should be suppressed in the production of
the $\varphi$ and $J/\psi$ mesons, but given the
present crudeness of the state of the art, we added the $f$
meson contribution in the $\varphi$ meson case. While we expect the $f$
contribution to $J/\psi$ meson production to be essentially
zero, we believe that it is
not irrelevant for $\varphi$ meson production due to $\omega - \phi$ mixing.
Indeed, in the $\varphi$ decay mode, more
than $15$\% is due to non strange particles and the $\bar K K$
decay mode is present in $f$ meson decay.

For $\omega$ meson photoproduction, we include also $\pi$ meson exchange
(see also the discussion in \cite{ref:DL}), which is needed in the
 low energy sector given that we try to describe the data for all
energies $W$. Even though we did not expect it,
the model 
describes well the data down to threshold.

In the integrated elastic cross section \be \displaystyle
\sigma(z, M_V^2, \tilde Q^2)^{\gamma p\rightarrow Vp}_{el} =
4\pi\int\limits_{t_{-}}^{t_{+}}dt|A^{\gamma p\rightarrow
Vp}(z,t;M_V^2,\tilde Q^2)|^{2}\; , \label{eq:sigma} \ee
 the upper and lower limits
 \be
2t_{\pm}=\pm
\frac{L_{1}L_{2}}{W^{2}}-(W^{2}+Q^{2}-M_{V}^{2}-2M_{p}^{2})+
\frac{(Q^{2}+M_{p}^{2})(M_{V}^{2}-M_{p}^{2})}{W^{2}}, \ee \be
L_{1}=\lambda(W^{2},-Q^{2},M_{p}^{2}),\qquad
L_{2}=\lambda(W^{2},M_{V}^{2},M_{p}^{2}), \ee \be
\lambda^{2}(x,y,z)=x^{2}+y^{2}+z^{2}-2xy-2yz-2zx, \ee are
determined by the kinematical condition $-1\leq
\cos\theta_{s}\leq 1$ where $\theta_{s}$ is the scattering angle
in the s-channel of the process.

The accurate account of the kinematically available $t$ region
allows us to describe effectively the threshold behaviour of
cross sections, so that when $W\rightarrow W_{threshold}$ we
have $t_{-}\rightarrow t_{+}$ and the elastic cross section goes
to zero. The imaginary part of the amplitude does not vanish at
threshold, but it turns out that the kinematical
cancellation alone accounts for the threshold behaviour.
The kinematical character of the threshold behaviour of the
integrated cross sections was studied long ago \cite{threshold}.

For the Pomeron contribution (\ref{eq:pomeron}) we use a
nonlinear trajectory \be\label{eq:trajectory_of_pomeron}
\alpha_\Pom (t)=1+\gamma (\sqrt{4m_\pi^2}-\sqrt{4m_\pi^2-t}\:),
\ee where $m_\pi$ is the pion mass. Such a trajectory was
utilized for photoproduction amplitudes in
\cite{ref:dipole_vector}, \cite{ref:Jenk} and its roots are very old
\cite{ref:Enrico1965}.

For the $f$-meson contribution for the sake of simplicity we use
the standard linear Reggeon trajectory \be \alpha_\Reg (t)=\alpha_\Reg
(0)+\alpha'_\Reg (0)\: t \: . \ee 

In the case of nonzero virtuality of the photon, we have a new
variable in play $Q^2=-q^2$. At the same time, the cross section 
$\sigma_L$ is nonzero. According to
\cite{ref:Brodsky}, QCD predicts the following dependence for
$\sigma_T$, $\sigma_L$ and their ratio as $Q^2$ goes to
infinity: \bea \nonumber
\sigma_T \sim \frac{1}{Q^8}(x_\Pom G(x_\Pom, \tilde Q^2/4))^2 ; \\
\label{eq:qcd}
\sigma_L \sim \frac{1}{Q^6}(x_\Pom G(x_\Pom, \tilde Q^2/4))^2 ; \\
\nonumber
R\equiv \sigma_L/\sigma_T \sim Q^2/M_V^2; \\
\nonumber
\sigma = (\sigma_T + \sigma_L) \Big|_{Q^2 \rightarrow \infty} \sim \sigma_L.
\nonumber
\eea
Wheere $x_\Pom G(x_\Pom, \tilde Q^2/4)$ is the gluon distribution function
and $x_\Pom=\frac{Q^2+M_V^2}{W^2+M_V^2}$ (see however \cite{ref:Cudell, ref:Cudell1} where another
possibility is investigated).

\section{The Model}
For the Pomeron resides we use the following parametrization
\bea\label{eq:couplings} g_i(t; M_V^2,\tilde
Q^2)=\frac{g_i}{Q_i^2+\tilde Q^2}
exp(b_i(t; \tilde Q^2))\; , \\
\nonumber
i=0,1 \: .
\nonumber
\eea
\noindent The slopes are chosen as
\bea\label{eq:slopes}
b_i(t; \tilde Q^2)=\Bigl(b_{i0}+\frac{b_{i1}}{1+\tilde Q^2/Q_{b}^2}\Bigr)(\sqrt{4m_\pi^2}-\sqrt{4m_\pi^2-t}\:)\; , \\
\nonumber i=0,1 \: , \eea to comply with the previous choice
(\ref{eq:trajectory_of_pomeron}) and analyticity requirements
\cite{ref:Enrico1965}.

\noindent The Reggeon residue is \be\label{eq:couplingsR}
g_\Reg(t; M_V^2, \tilde Q^2)=\frac{g_\Reg
M_p^2}{(Q_\Reg^2+\tilde Q^2)\tilde Q^2}exp(b_\Reg(t; \tilde
Q^2))\; , \ee where \be\label{eq:slopesR} b_\Reg(t; \tilde
Q^2)=\frac{b_\Reg}{1+\tilde Q^2/Q_{b}^2}t\; , \ee $g_0, \; g_1$,
$Q_0^2\; (GeV^2)$, $ Q_1^2\; (GeV^2)$, $ Q_\Reg^2\; (GeV^2)$, $
Q_b^2\; (GeV^2)$, $b_{00}\; (GeV^{-1})$, $b_{01}\; (GeV^{-1})$,
$b_{10}\; (GeV^{-1})$, $b_{11}\; (GeV^{-1})$, $b_{\Reg}\;
(GeV^{-2})$ are adjustable parameters. $\Reg=f$ for $\rho$ and
$\varphi$, $\Reg=f,\pi$ for $\omega$. We use the same slope
$b_\Reg$ for $f$ and $\pi$ Reggeon exchanges.

\subsection{Photoproduction of vector mesons by real photons ($Q^2=0$).}

In the fit we use all available data starting from the threshold
for each meson. As the new dataset of ZEUS \cite{NEWZEUS} provides us with the
unique information on both integrated elastic cross section
and differential distribution of exclusive $J/\psi$ meson photoproduction,
we keep only these data for $Q^2=0$. This allows us to explore
the effects of nonlinearity of the Pomeron trajectory and residues.
In the region of non zero $Q^2$ the combined
data of H1 and ZEUS is used.

Different experiments have different normalization especially at
low energies. This implies that the $\chi^2/{\rm d.o.f.}$ will
not be very good while the overall agreement is quite
satisfactory.

The whole  set of data is composed of $357$  experimental points
\footnote{The data are available at \\
REACTION DATA Database {\it http://durpdg.dur.ac.uk/hepdata/reac.html} \\
CROSS SECTIONS PPDS database {\it http://wwwppds.ihep.su:8001/c1-5A.html}}
 and, with
a grand total of $12$ parameters, we find $\chi^2/{\rm d.o.f}=1.49$. The
main contribution to $\chi^2$ comes from the low energy region ( $W\le
4\;GeV$); had we started fitting from $W_{min}=4\;GeV$, the resulting
$\chi^2/{\rm d.o.f}=0.85$ for the elastic cross sections  would be much better and more appropriate for a
high energy model.

In order to get a reliable description and the parameters
of the trajectories and residues we use elastic cross sections
for each process from threshold  up to the highest
values of the energy and  differential cross sections in the whole t-region 
where data are available: 0 $\le |t| \le$ 1.6 $GeV^2$. The data
on differential cross section of $\rho$ meson production at $W=71.3\; GeV$ and $\varphi$ meson
production at $W=13.731\; GeV$ are not included in the fitting procedure.

The parameters are given in Table \ref{Table 1.}. The errors on the
parameters are obtained by MINUIT.

{\small
\begin{table}[H]
\begin{center}
\begin{tabular}{|l|l|r|r||l|l|lr|r|}
\hline
N & Parameter & Value & Error & & Trajectory & &$\alpha (0)$ & $\alpha' (0)\; GeV^{-2}$\\
\hline
1 & $\gamma \: (GeV^{-1})$  & 0.53853E-01 &  0.15666E-01& &     && FIXED & FIXED  \\
2 & $g_1$                   & 0.10435E-01 &  0.17851E-03& 1&$f$ Reggeon   && 0.8 &  0.85 \\
3 & $g_0$                   &-0.32901E-01 &  0.49449E-04& 2&$\pi$ Reggeon && 0.0 &  0.85 \\
\hhline{|~|~|~|~||=|=|==|=|}
4 & $g_f$                   & 0.83371E-01 &  0.49503E-03& & Meson & \# &of points & $\chi^2$ per point\\
\cline{5-9}
5 & $g_\pi $                & 0.60011     &  0.21962E-01& 1 & $\rho_0(770)$&$\sigma_{el}$,&  127 & 1.49 \\
6 & $Q_0^2\; (GeV^2)$       & 0.0         &       FIXED&   & $\rho_0(770)$&$\frac{d\sigma_{el}}{dt}$,&  24 & 0.99 \\
7 & $Q_1^2\; (GeV^2)$       & 0.41908     &  0.23586E-02& 2 & $\omega(782)$& $\sigma_{el}$,&57  & 1.65 \\
8 & $Q_\Reg^2\; (GeV^2)$    & 0.0         &       FIXED&   & $\omega(782)$& $\frac{d\sigma_{el}}{dt}$,& 12 & 0.83 \\
8 & $Q_b^2\; (GeV^2)$       & 3.9724      &  0.32482& 3 & $\varphi(1020)$ & $\sigma_{el}$,& 39  & 0.98 \\
10 & $b_{10}\; (GeV^{-1})$  & 2.1251      & 0.73983E-01&   & $\varphi(1020)$ & $\frac{d\sigma_{el}}{dt}$,& 5  & 0.61 \\
11 & $b_{11}\; (GeV^{-1})$  & 2.5979      & 0.21451& 4 & $J/\psi(3096)$&$\sigma_{el}$,& 29  & 0.79 \\
12 & $b_{00}\; (GeV^{-1})$  & 2.6967      & 0.24985E-01&   & $J/\psi(3096)$&$\frac{d\sigma_{el}}{dt}$,& 70  & 1.92 \\
\hhline{|~|~|~|~||=|=|==|=|}
13 & $b_{01}\; (GeV^{-1})$  & 6.7897      & 0.18717E-01& & All mesons & \# &of points & $\chi^2/{\rm d.o.f.}$  \\
\cline{5-9}
14 & $b_{\Reg}\; (GeV^{-2})$ & 4.5741     &  0.10509E-02&&$\rho_0,\omega,\varphi,J/\psi $ && 357 & 1.49 \\
\hline
\end{tabular}
\end{center}
\vskip -0.5cm \caption{ Parameters obtained by fitting
$\rho_0,\omega,\varphi$ and $J/\psi$ photoproduction data
\label{Table 1.}}
\end{table}}

\begin{figure}[H]
\centering {\vspace*{ -1cm} \epsfxsize=130mm
\epsffile{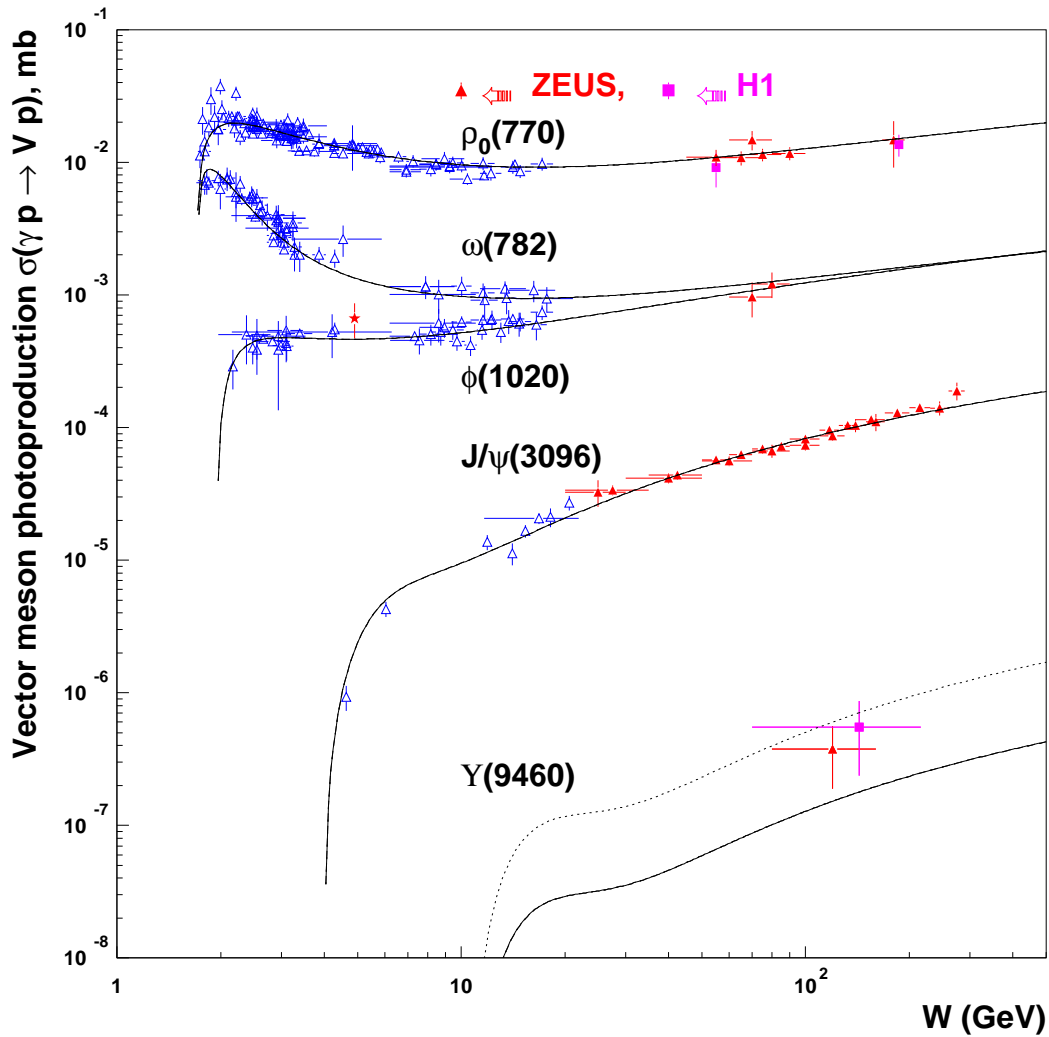}} \vskip-2.cm \caption{Elastic
cross-sections for vector meson photoproduction. The solid curve for
$\Upsilon (9460)$ production corresponds to
$N_\Upsilon=N_\varphi$, the dotted line to $N_\Upsilon=N_{J/\psi}$.
\label{fig:mesons}}
\end{figure}

The results are presented in Fig. \ref{fig:mesons}, which shows also the
 prediction of the model for $\Upsilon (9460)$ photoproduction.

As can be seen, the model describes the vector meson
exclusive photoproduction data without the need of Pomeron
contribution with intercept higher than 1. In addition, the
rapid rise of the $J/\psi$ cross section at low energies is
described as a transition phenomenon, a delay of the onset of
the real asymptotic.

\noindent Had one assumed SU(5) flavour symmetry for the
$\Upsilon (9460)$, we would have found $<Q^2_j>_\Upsilon=1/9$
and thus $N_\Upsilon=N_\varphi$. This relation leads to
underestimate the $\Upsilon$ photoproduction cross section (see
solid line Fig. \ref{fig:mesons}). Phenomenologically we find that
$N_\Upsilon=N_{J/\psi}$ gives a better description of  the data on
$\Upsilon (9460)$ production (dotted line), but perhaps an intermediate value
would be more appropriate.

\subsection{Differential cross section of vector meson exclusive production}

\begin{figure}[H]
\parbox[c]{8.7cm}{\epsfxsize=76mm
\epsffile{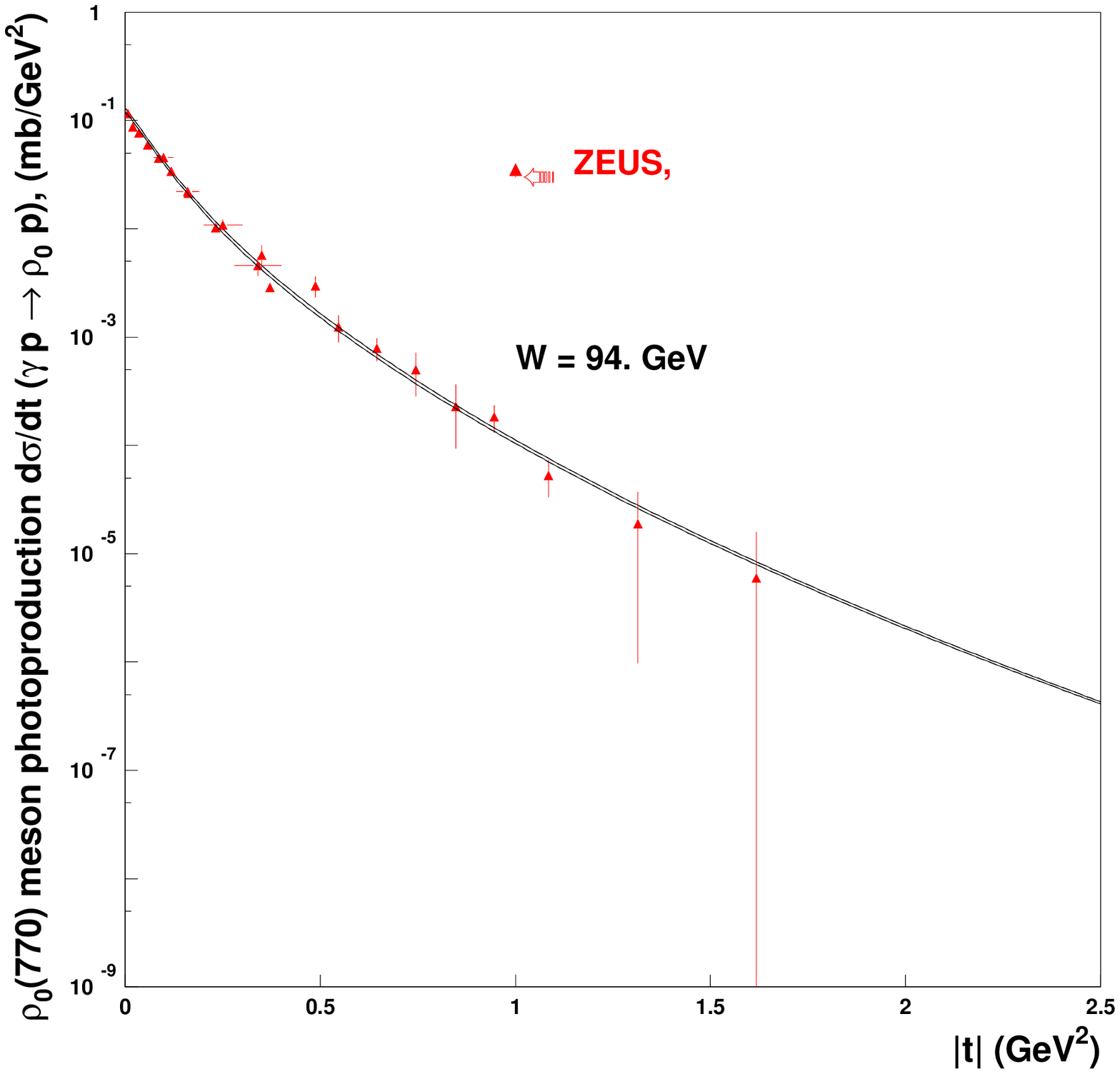}} \hfill~\parbox[c]{7.6cm}{\epsfxsize=76mm
\epsffile{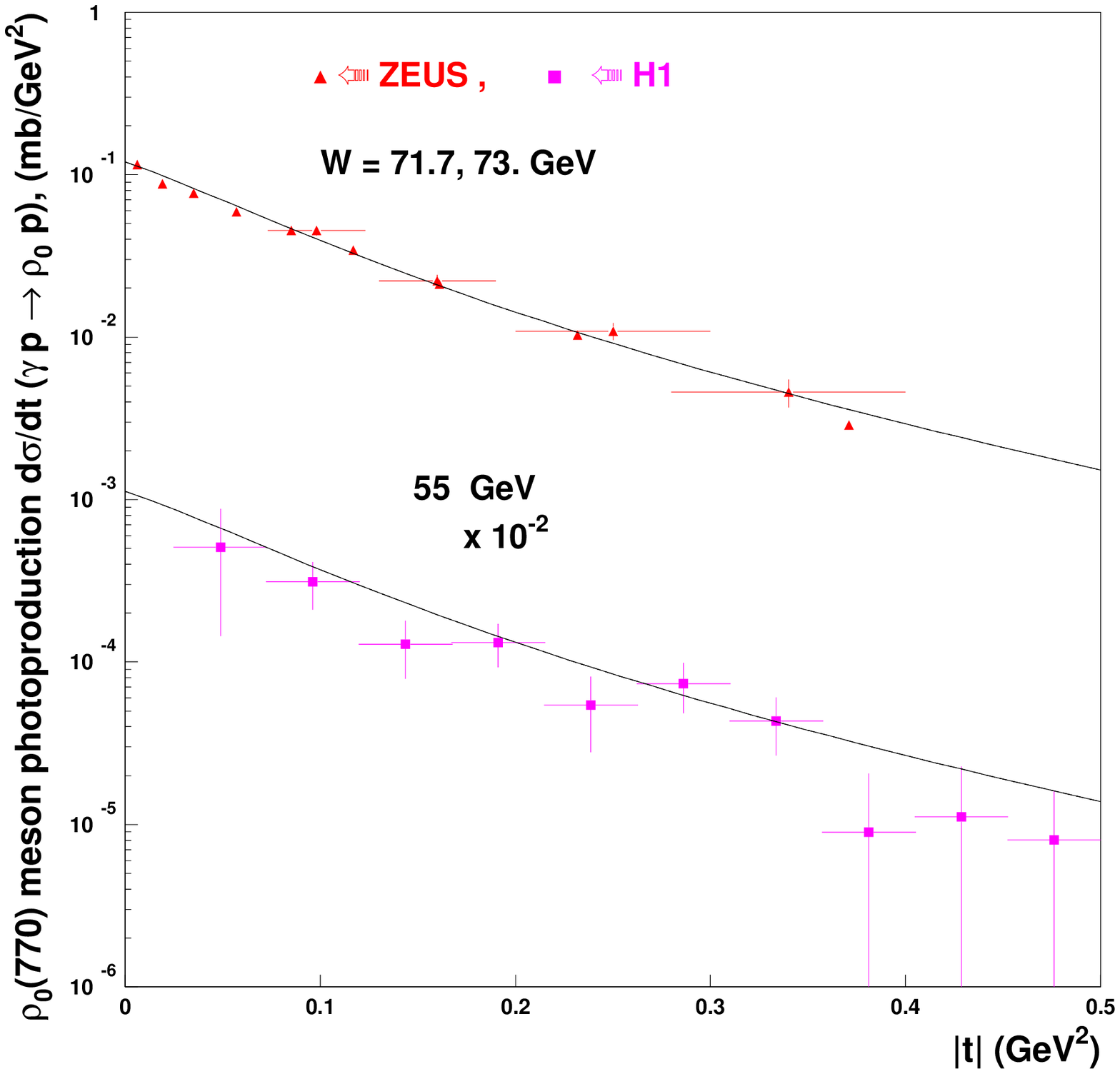}}

\vspace*{-1.3cm}
\parbox[t]{7.7cm}{\caption{Differential cross section of exclusive
$\rho_0$ meson photoproduction for $W=94\; GeV$.\label{fig:rhod}}}
\hfill~\parbox[t]{7.7cm}{\caption{Differential cross section of exclusive
$\rho_0$ meson photoproduction for $W=71.7,\; 73,\; {\rm and}\; 55\; GeV$.
The data and curves for $W=55\; GeV$ are scaled by a factor $10^{-2}$.
\label{fig:rhod1}}}
\end{figure}

\begin{figure}[H]
\centering {\vspace*{ -1cm} \epsfysize=120mm \epsffile{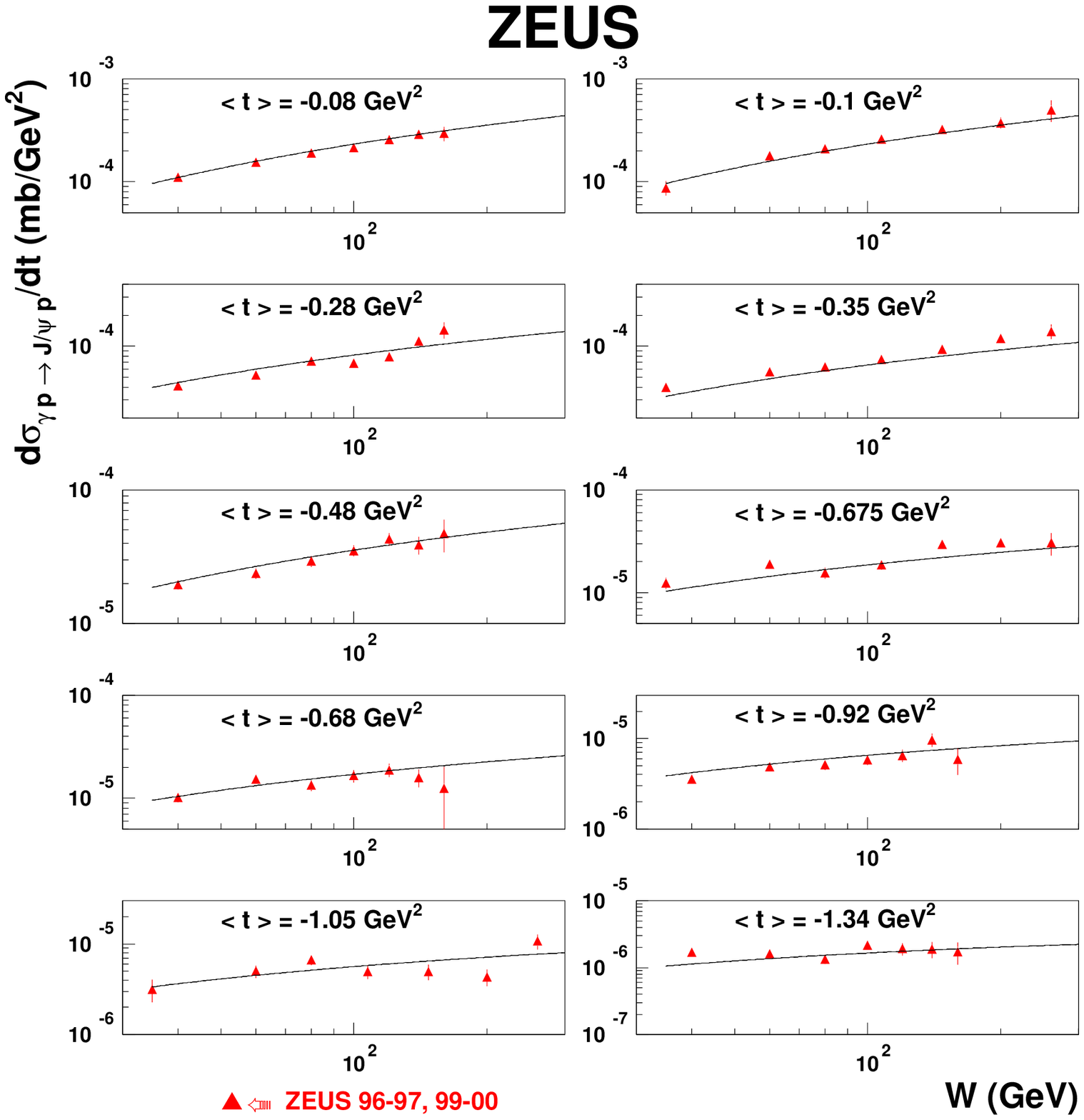}} \vskip
-1.9cm \caption{Differential cross section of exclusive $J/\psi$ meson
photoproduction as a function of $W$ at different $<t>$.
\label{fig:jpsidt} }
\end{figure}

\begin{figure}[H]
\centering {\vspace*{ -1.5cm} \epsfysize=120mm \epsffile{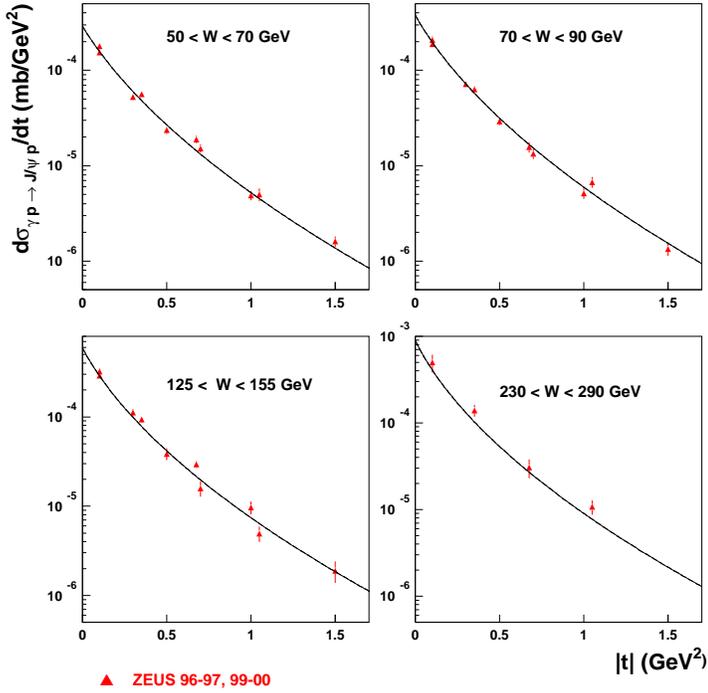}} \vskip
-1.9cm \caption{Differential cross section of exclusive $J/\psi$ meson
photoproduction.
\label{fig:jpsid} }
\end{figure}

\begin{figure}[H]
\parbox[c]{8.7cm}{\epsfxsize=76mm
\epsffile{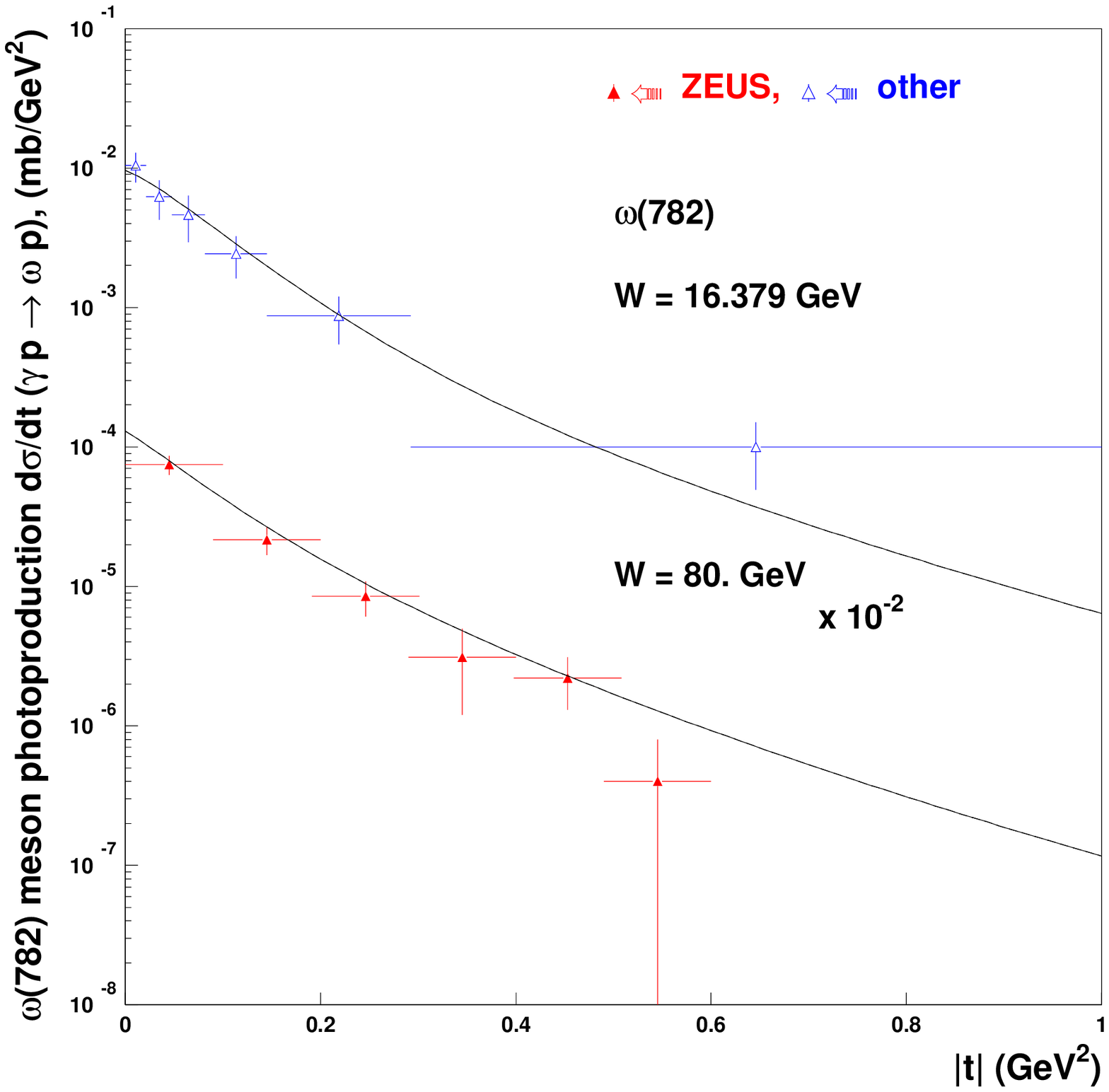}} \hfill~\parbox[c]{7.6cm}{\epsfxsize=76mm
\epsffile{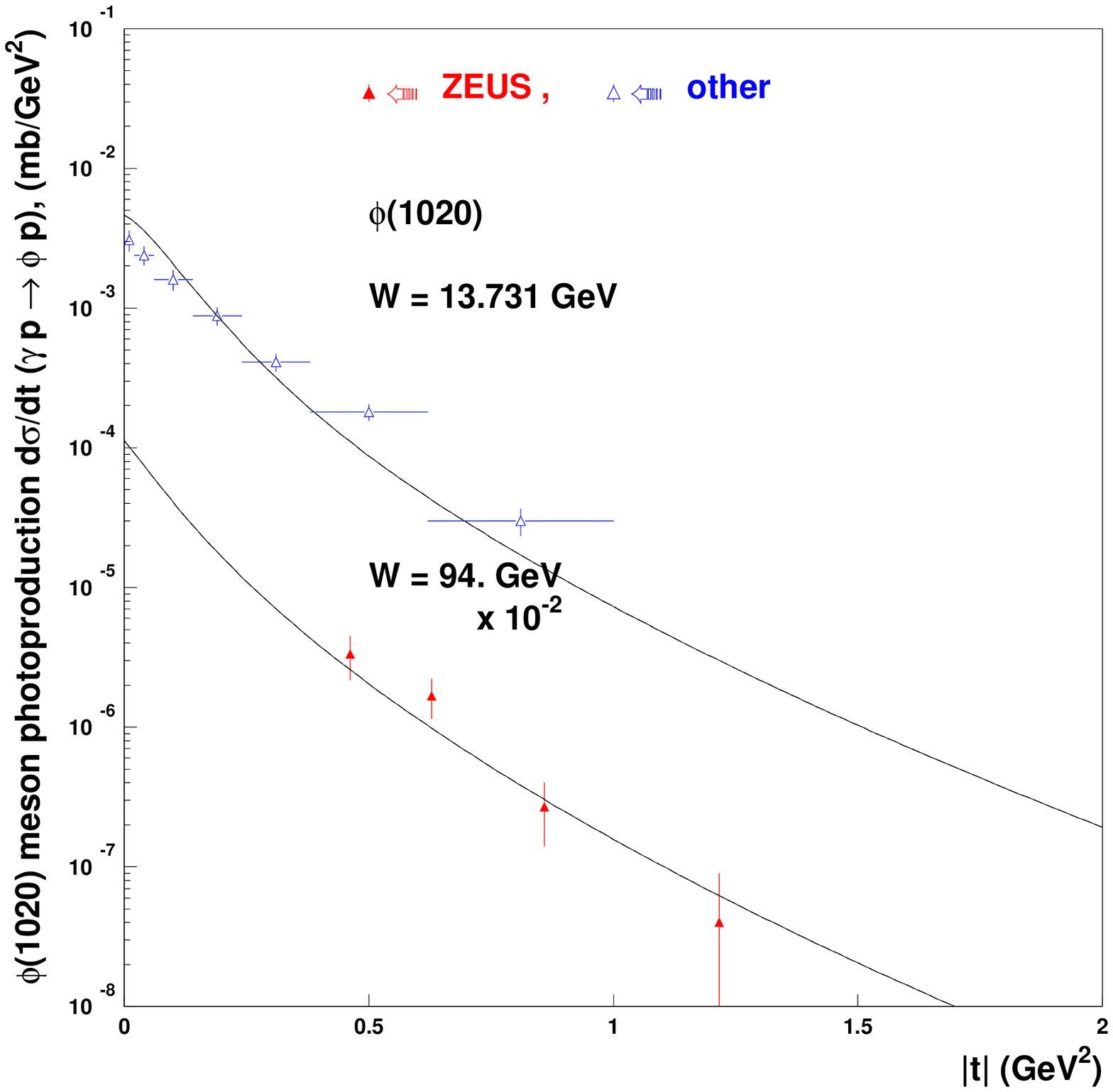}}

\vspace*{-1.7cm}
\parbox[t]{7.7cm}{\caption{Differential cross section of exclusive
$\omega$ meson photoproduction for $W=16.379\;{\rm and}\;80\; GeV$. The
data and curves for $W=80\; GeV$ are scaled by a factor $10^{-2}$.
\label{fig:omegad}}}
\hfill~\parbox[t]{7.7cm}{\caption{Differential cross section of
exclusive $\varphi$ meson photoproduction for $W=13.371\;{\rm and}\; 94\;
GeV$. The data and curves for $W=94\; GeV$ are scaled by a factor $10^{-2}$.
\label{fig:phid}}}
\end{figure}

The differential cross section is given by
\be
\frac{d\sigma}{dt} = 4\pi |A(z,t;\tilde Q^2, M_V^2)|^2\; .
\ee
Using the amplitude from the previous section this quantity is now
calculated and the comparison with the data is presented in Fig.
\ref{fig:rhod}, \ref{fig:rhod1}, \ref{fig:jpsid}, \ref{fig:jpsidt},
\ref{fig:omegad} and \ref{fig:phid}.

Given the universality of our approach we conclude 
that extracting the Pomeron trajectory from the
experimental data as proposed in \cite{ref:H1JPSI} and \cite{NEWZEUS} using the data depicted
in Fig.~\ref{fig:jpsidt} cannot be regarded as a valid argument to support 
either hard Pomeron contribution or the BFKL Pomeron.

\vspace*{ -1cm}

Exploring the nonlinearity of the Pomeron trajectory (16) and
slopes (19), we have tried adding to either a linear term or
a heavier threshold; both give
negligibly small effects. Thus we conclude that new ZEUS data on
$d\sigma_{J/\psi}/dt$ (see for example Fig. \ref{fig:jpsid}) are a
strong support in favor of the nonlinear Pomeron trajectory.


\subsection{Photoproduction of vector mesons by virtual photons\\ ($Q^2>0$).}

In (\ref{eq:couplings}) and (\ref{eq:couplingsR}) the $Q^2$-
dependence ($\tilde Q^2=Q^2+M_V^2$) is completely fixed up to an
{\it a priori} arbitrary dimensionless function $f(Q^2)$ such
that $f(0)=1$. Thus, we may introduce a new factor that
differentiates virtual from real photoproduction:

\be\label{eq:fpom}
f(Q^2)=\Big(\frac{M_V^2}{\tilde Q^2}\Big)^n
\ee

Accordingly, in the case $Q^2\neq 0$ we use the following parametrizations
for Pomeron couplings (compare with Eq. \ref{eq:couplings}):
\be\label{eq:couplingsqP}
\hat g_i(t; \tilde Q^2, M_V^2)= f(Q^2)g_i(t; \tilde Q^2, M_V^2),\; i=0,1,
\ee
where, for the sake of completeness, we will examine three different
{\it choices} for the asymptotic $Q^2$ behaviour of the Pomeron
residue

\noindent
{\it \underline{Choice I}}
 \be\label{eq:couplingsChoice}
n=1, \qquad
 \sigma_T({Q^2 \rightarrow \infty}) \sim \frac{1}{Q^8}\; .
\ee {\it \underline{Choice II}}
 \be \label{eq:couplings1}
n=0.5, \qquad
 \sigma_T({Q^2 \rightarrow \infty}) \sim \frac{1}{Q^6}\; .
\ee
{\it \underline{Choice III}}
 \be \label{eq:couplings2}
n=0.25, \qquad
 \sigma_T({Q^2 \rightarrow \infty}) \sim \frac{1}{Q^5}\; .
\ee

\noindent For the reggeon couplings we have
\be
f_\Reg(Q^2)=\Big(\frac{c_1 M_V^2}{c_1 M_V^2+Q^2}\Big)^{n_2}\; ,
\ee
where $c_1$ is an adjustable parameter and
$n_2=0.25, \; -0.25, \; -0.5$ for {\it choice I, II, III}.

Accordingly, in the case $Q^2\neq 0$ we use the following parametrizations
for Reggeons couplings (compare with Eq. \ref{eq:couplingsR}):
\be\label{eq:couplingsq}
\hat g_\Reg(t; \tilde Q^2, M_V^2)= f_\Reg(Q^2)g_\Reg(t; \tilde Q^2, M_V^2)\; .
\ee

The lack of data on the ratio $\sigma_L/\sigma_T$, especially in
the high $Q^2$ domain, does not allow us to draw definite
conclusions about its asymptotic behaviour (the Regge theory is
not the appropriate tool for giving predictions in this case),
nor do we have a unique prescription in the framework of our
model. There may be several realizations of the model with
different asymptotic behaviour of $\sigma_L/\sigma_T$
\cite{owrmodel}. As a demonstration of such a possibility we
explore the predictions (\ref{eq:qcd}) and use the following
(most economical) parametrization for $R$ (which cannot be
deduced from the Regge theory)

\noindent{\it \underline{Choice I, II, III}}
\be R(Q^2, M_V^2) =
\Big(\frac{c M_V^2+Q^2}{c M_V^2}\Big)^{n_1}-1
\label{eq:ratio}
\ee
where $c$ and $n_1$
are adjustable parameters for {\it choice I, II, III}.

We have, thus, 3 additional adjustable parameters as compared with
real photoproduction. In order to obtain the values of
the parameters for the case $Q^2\ne 0$,
we fit just the data\footnote{The data are available at \\
REACTION DATA Database {\it http://durpdg.dur.ac.uk/hepdata/reac.html} \\
CROSS SECTIONS PPDS database {\it http://wwwppds.ihep.su:8001/c1-5A.html}}
on $\rho_0$ meson photoproduction in the region $0\le Q^2\le 35 \; GeV^2$;
the parameters for photoproduction by real photons are the same as in
Table \ref{Table 1.}. In order to avoid the low $W$ region where nucleon
resonances may spoil the picture of $\rho$ meson exclusive production, we
restrict ourselves to the energy domain $W\geq 4\; GeV$ for $Q^2 \ne 0$.

The parameters thus obtained are shown in Table \ref{Table 2.}.
\begin{table}[H]
\begin{center}
\begin{tabular}{|l|l|l|l|l|}
\hline
 &  &{\it \underline{Choice I}} & {\it \underline{Choice II}}& {\it \underline{Choice III}}\\
\hline
N & Parameter & Value & Value & Value\\
\hline
1 & $c$ & 1.2666    $\pm$   0.048& 1.6900    $\pm$   0.167 &3.3282    $\pm$   0.916\\
2 & $n_1$ & 1.8355  $\pm$     0.026& 0.84596   $\pm$    0.033 & 0.32453   $\pm$    0.043\\
3 & $c_1$ & 2.3258  $\pm$     0.286  & 0.55469  $\pm$   0.044 & 0.78464  $\pm$     0.028\\
\hline
\hline
& Fit, \# of points &  $\chi^2/{\rm d.o.f.}$ &$\chi^2/{\rm d.o.f.}$ &  $\chi^2/{\rm d.o.f.}$ \\
\hline
&$\rho_0(770)$,  283 &  1.47& 1.53 & 1.56\\
\hline
\hline
& Meson,  \# of points &  $\chi^2$ per point & $\chi^2$ per point&  $\chi^2$ per point \\
\hline
1 & $\rho_0$,  283&   1.45 & 1.51 & 1.54\\
2 & $\omega$,  67&   1.46 {\bf (no fit!)}& 1.46 {\bf (no fit!)} & 1.46 {\bf (no fit!)}\\
3 & $\varphi$,  56&   0.78 {\bf (no fit!)}& 0.79 {\bf (no fit!)}& 0.82 {\bf (no fit!)}\\
2 & $J/\psi$, 54&   0.89 {\bf (no fit!)}& 0.92 {\bf (no fit!)}& 0.99{\bf (no fit!)}\\
\hline
\end{tabular}
\end{center}
\vskip -0.5cm
\caption{ Parameters obtained by fitting  $\rho_0$
virtual photoproduction data for {\it choice I, II, III.}\label{Table 2.}}
\end{table}

\begin{figure}[H]
\parbox[c]{8.6cm}{\epsfxsize=70mm
\epsffile{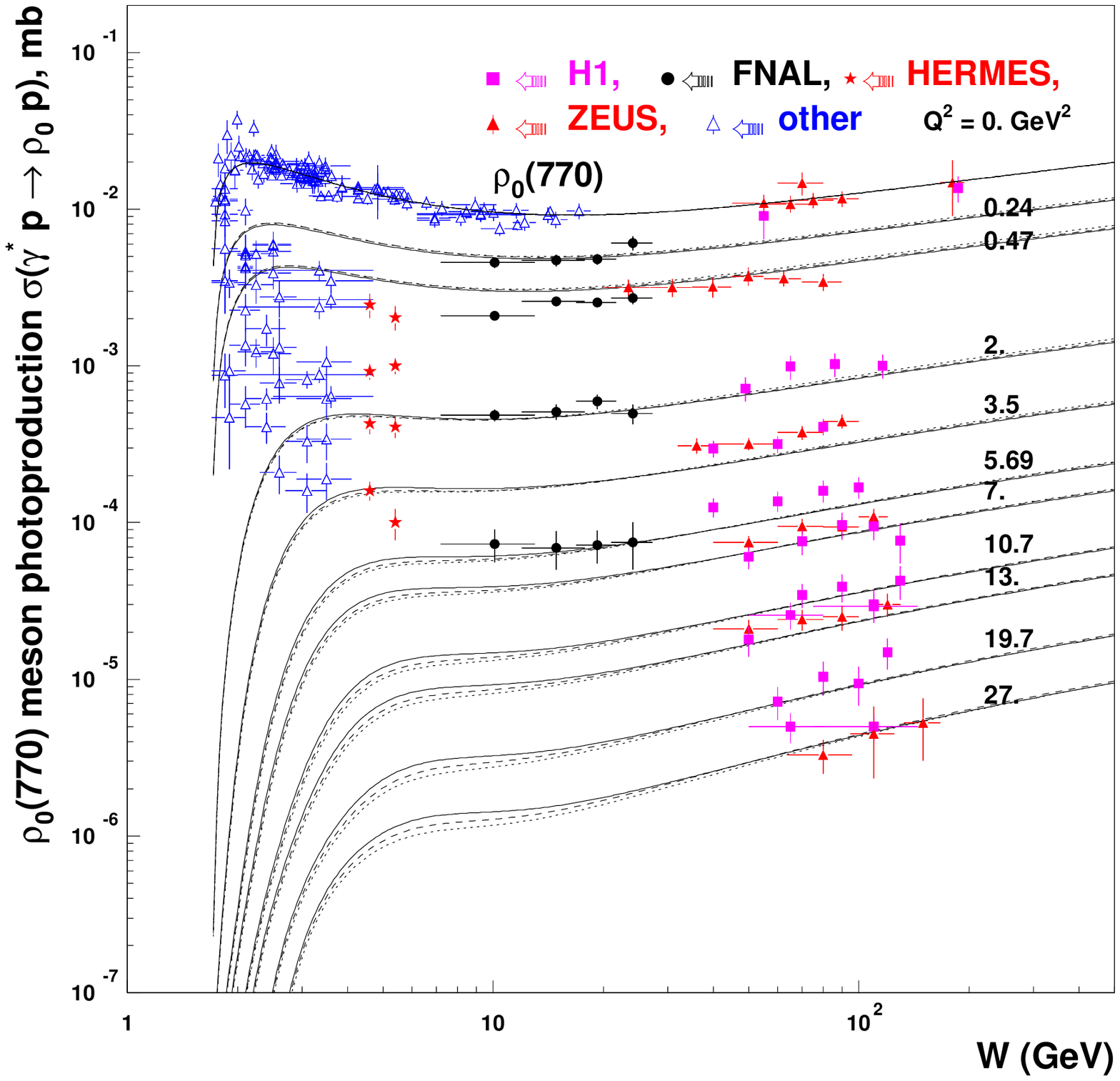}} \hfill~\parbox[c]{7.6cm}{\epsfxsize=70mm
\epsffile{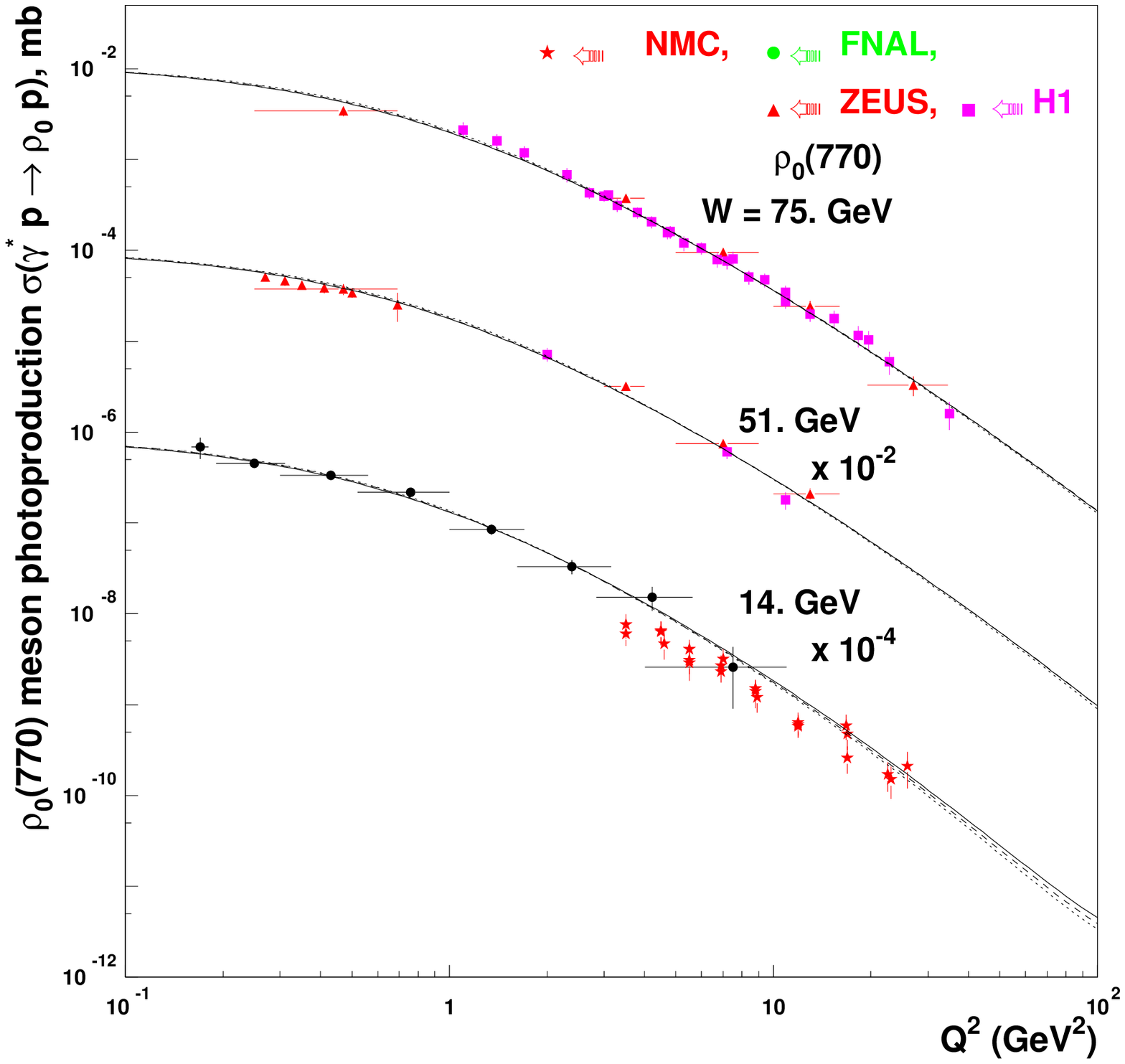}}

\vspace*{-1.3cm}
\parbox[t]{7.7cm}{\caption{Elastic cross section of exclusive
$\rho_0$ virtual photoproduction as a function
of $W$ for different values of $Q^2$. \label{fig:rho}}}
\hfill~\parbox[t]{7.7cm}{\caption{Elastic cross section of
exclusive $\rho_0$ virtual photoproduction as a
function of $Q^2$ for $W=75,\;51,\;{\rm and}\;14\; GeV$. The
data and curves for $W=51,\;{\rm and}\;14\; GeV$ are scaled by
factors $10^{-2}$ and $10^{-4}$. \label{fig:rhoq}}}
\end{figure}

\begin{figure}[H]
\parbox[c]{8.6cm}{\epsfxsize=70mm
\epsffile{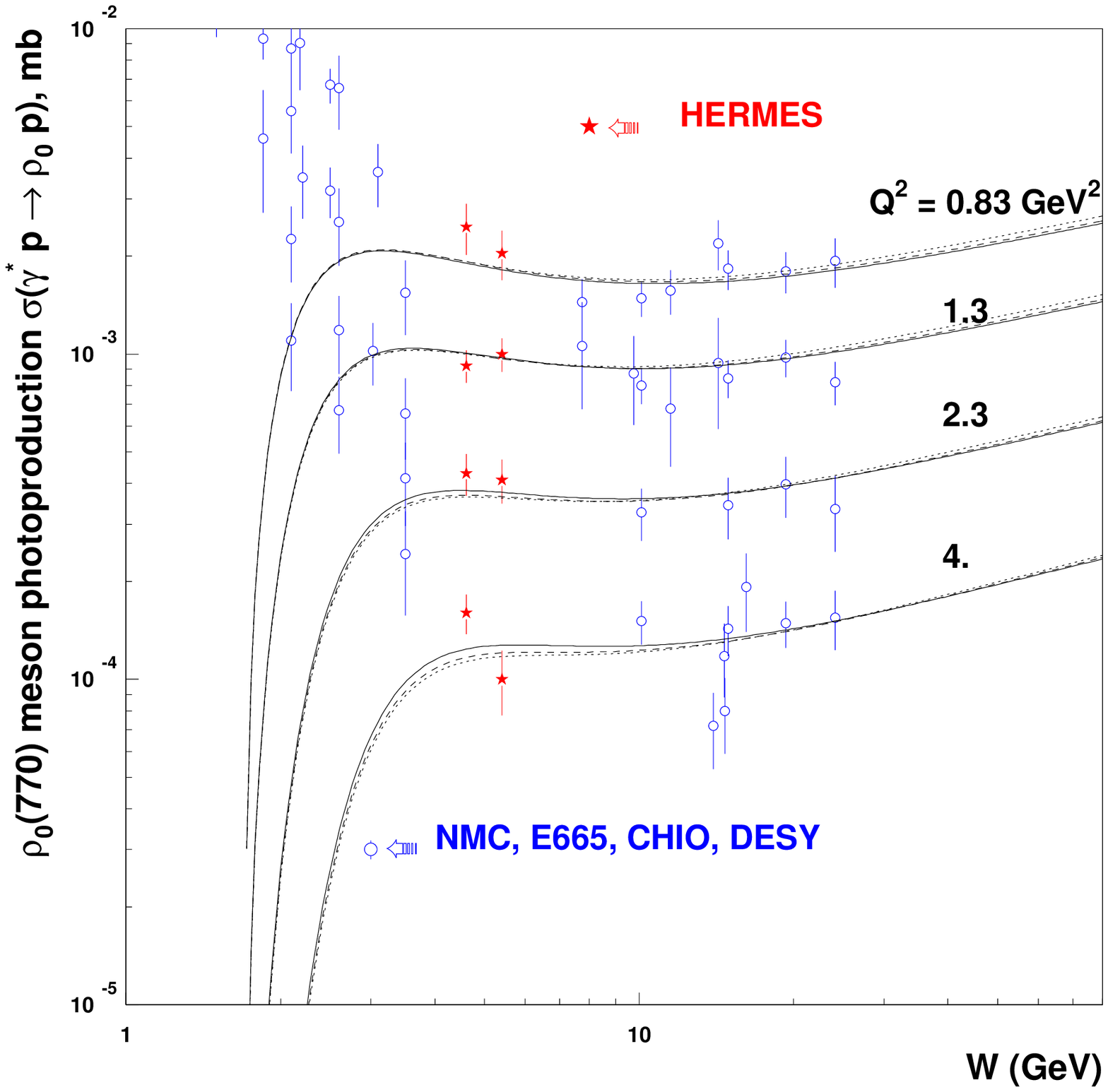}}
\hfill~\parbox[c]{7.6cm}{\epsfxsize=70mm \epsffile{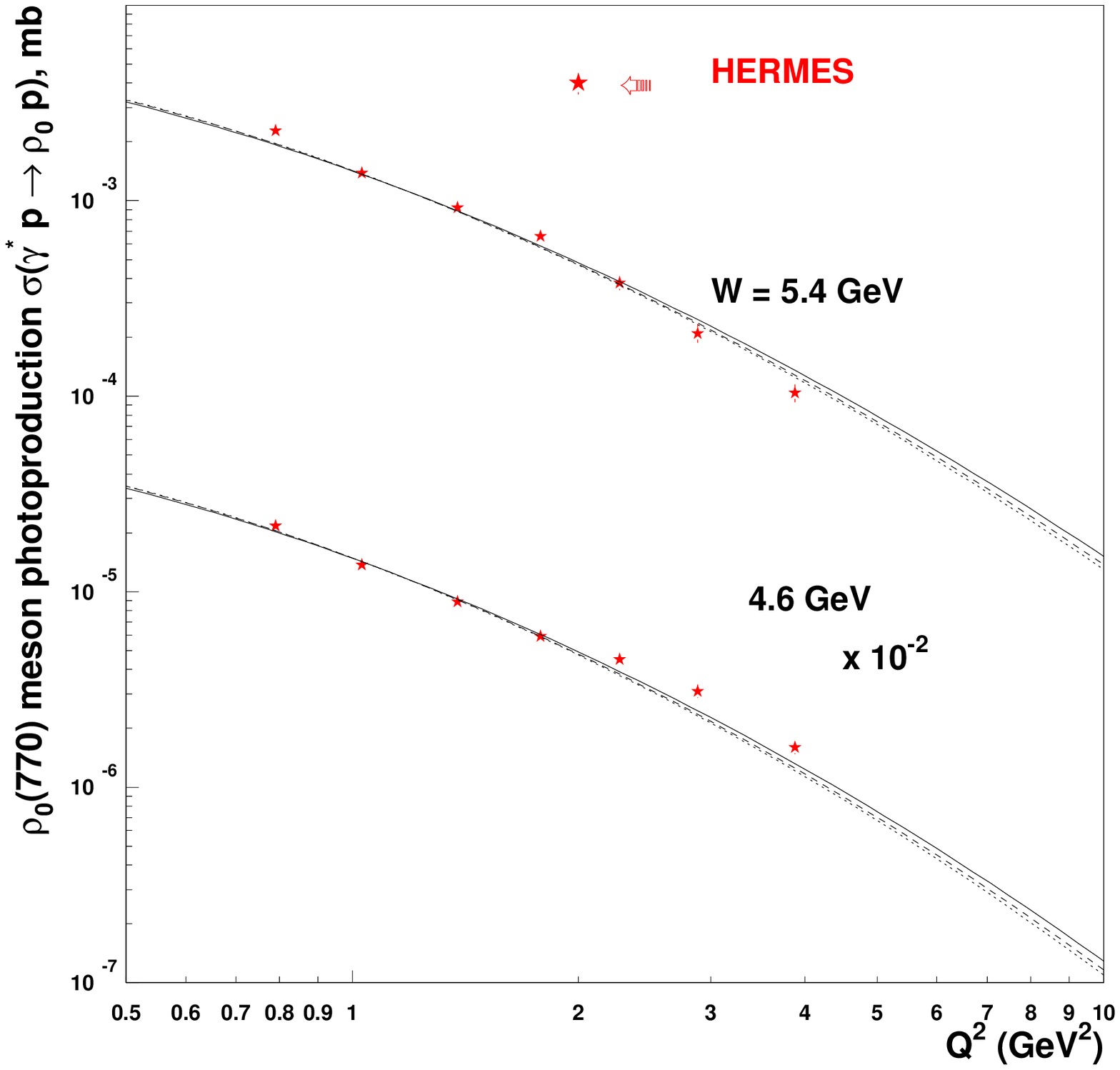}}

\vspace*{-1.3cm}
\parbox[t]{7.6cm}{\caption{Elastic cross section of exclusive $\rho_0$
virtual photoproduction as a function of $W$ for
various $Q^2$ in the region of low and intermediate $W$.
 \label{fig:rhohermes}}}
\hfill~\parbox[t]{7.6cm}{\caption{Elastic cross section of
exclusive $\rho_0$ virtual photoproduction as a
function of $Q^2$ for $W=5.4,\;{\rm and}\;4.6\; GeV$. The data
and curves for $W=4.6\; GeV$ are scaled by a factor $10^{-2}$.
\label{fig:rhoq1}}}
\end{figure}

The results of the fit are depicted in Figs. \ref{fig:rho},
\ref{fig:rhoq}, \ref{fig:rhohermes}, \ref{fig:rhoq1}. In these figures as well as in all following ones the solid lines,
dashed
lines and dotted lines correspond to the {\it choice I, II, III}
correspondingly.
 The description of
the data is very good at all energies. Both high energy data
from ZEUS and H1 Fig. \ref{fig:rho} and low energy data from HERMES Fig.
\ref{fig:rhohermes} are accounted for. In the region of the HERMES data
(Fig.~\ref{fig:rhohermes}) our description is comparable to the one of
Haakman, Kaidalov and Koch \cite{ref:Kaidalov} (see \cite{ref:Hermes} for
details).

We can now check the predictions of the model. As stated earlier, we aim
at a unified model for all vector meson production, thus the
only variable that changes is the mass of the vector meson. In the
following figures we depict our {\it predictions} for $\omega$, $\varphi$
and $J/\psi$ mesons and we compare them with the available data. The {\it
description} of the data is very good for all the three mesons. The
$\chi^2=0.89$ for $J/\psi$ meson exclusive production follows without any
fitting. Both $W$ and $Q^2$ dependences are reproduced very well. Notice that,
so far, the three {\it choices I, II, III} all give equally acceptable reproduction
of the data.

\begin{figure}[H]
\parbox[c]{8.7cm}{\epsfxsize=66mm
\epsffile{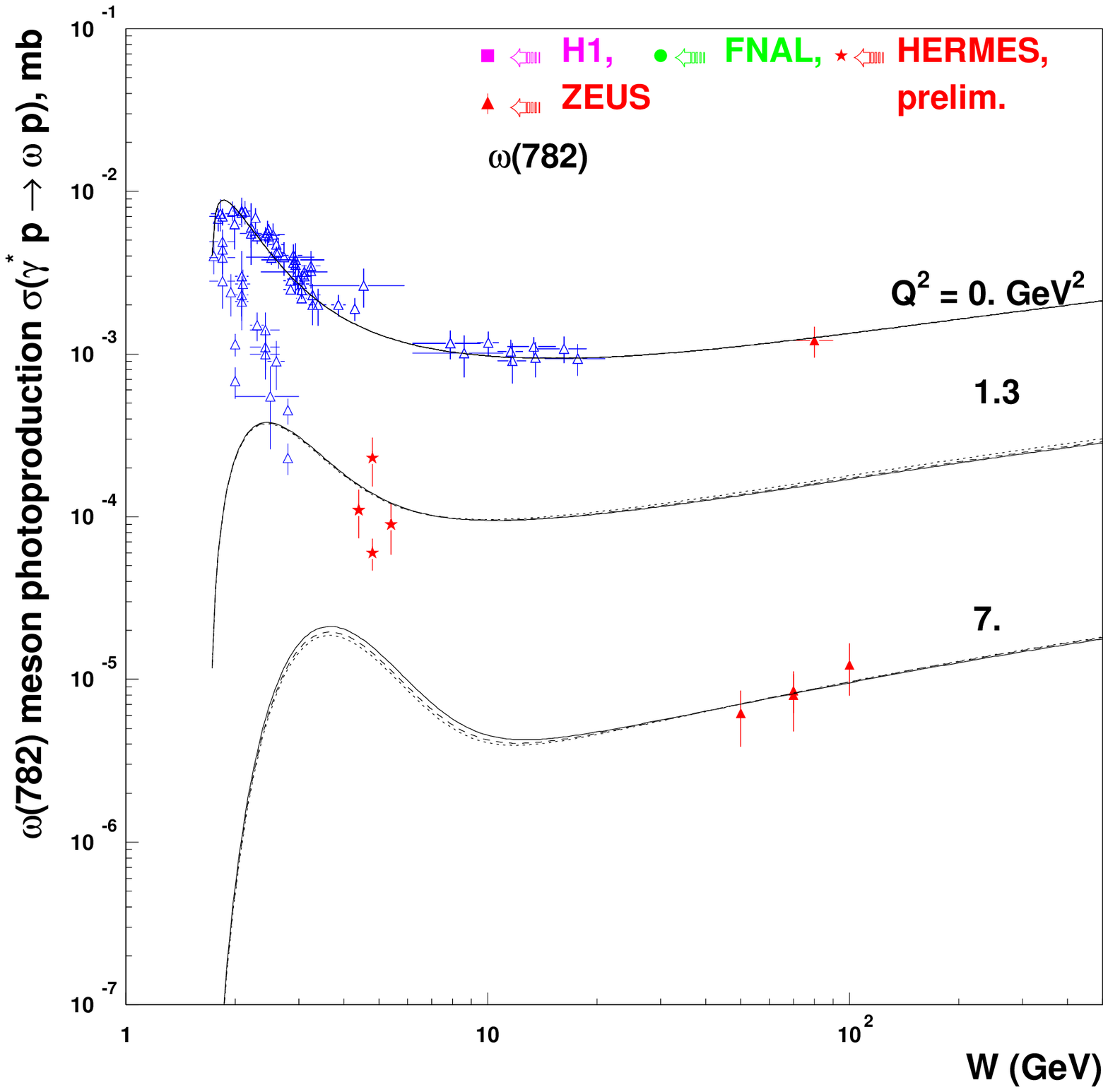}} \hfill~\parbox[c]{7.6cm}{\epsfxsize=66mm
\epsffile{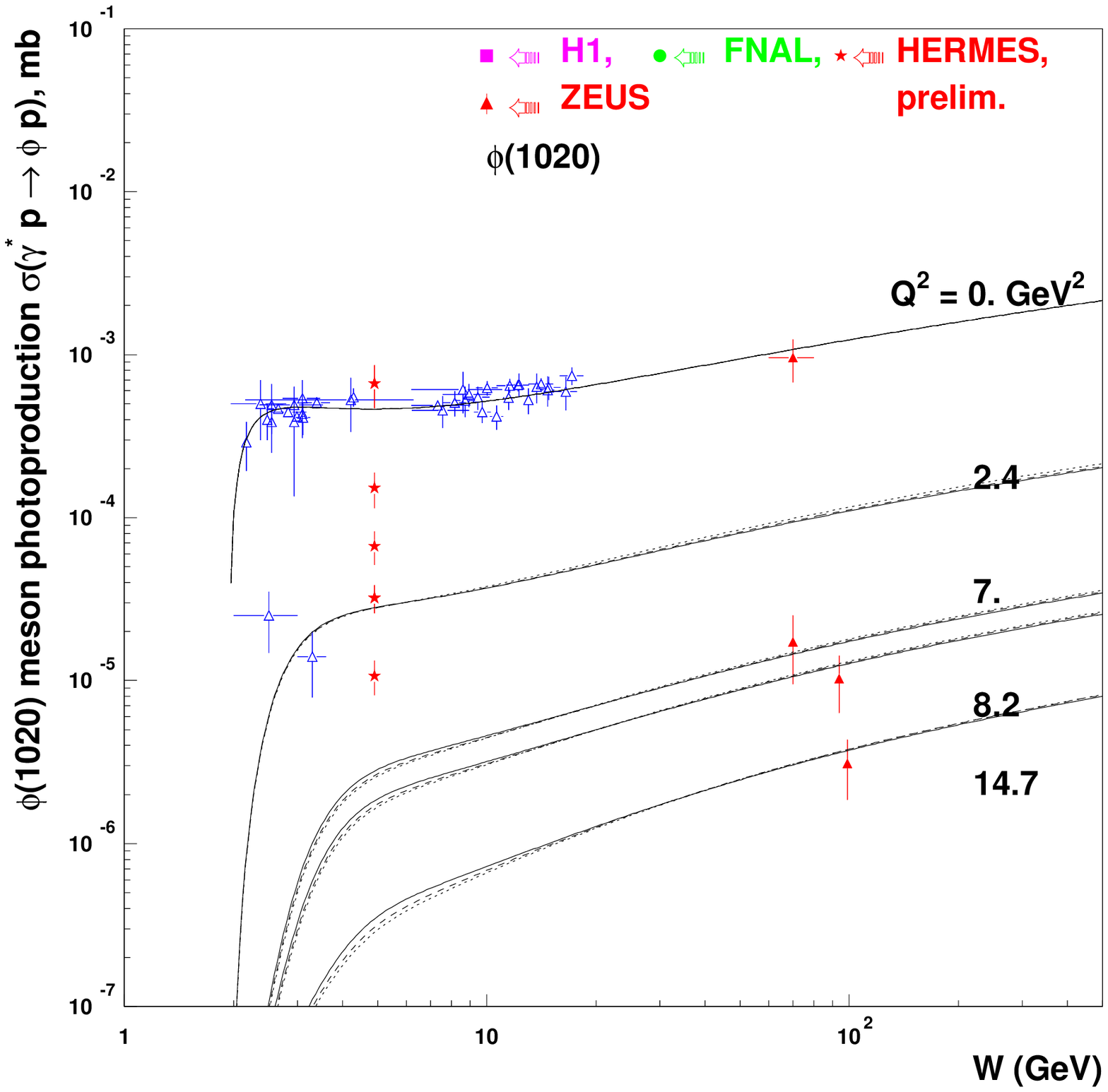}}

\vspace*{-1.6cm}
\parbox[t]{7.7cm}{\caption{Elastic cross section of exclusive $\omega$
virtual photoproduction as a function of $W$
for various $Q^2$. \label{fig:omega}}}
\hfill~\parbox[t]{7.7cm}{\caption{Elastic cross section of
exclusive $\varphi$ virtual photoproduction as
a function of $W$ for various $Q^2$.
\label{fig:phi}}}

\end{figure}

\begin{figure}[H]
\parbox[c]{8.7cm}{\epsfxsize=66mm
\epsffile{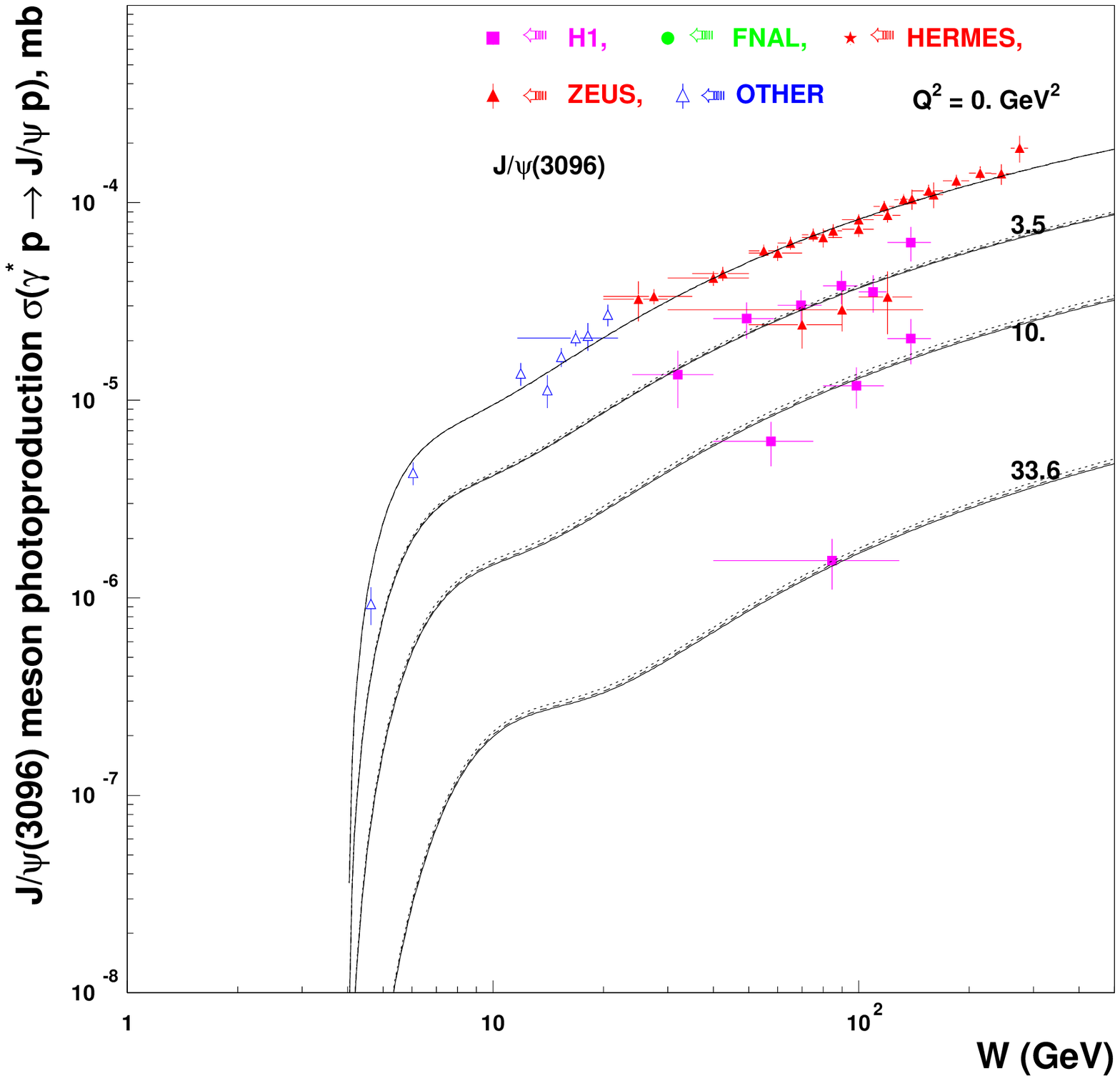}} \hfill~\parbox[c]{7.6cm}{\epsfxsize=66mm
\epsffile{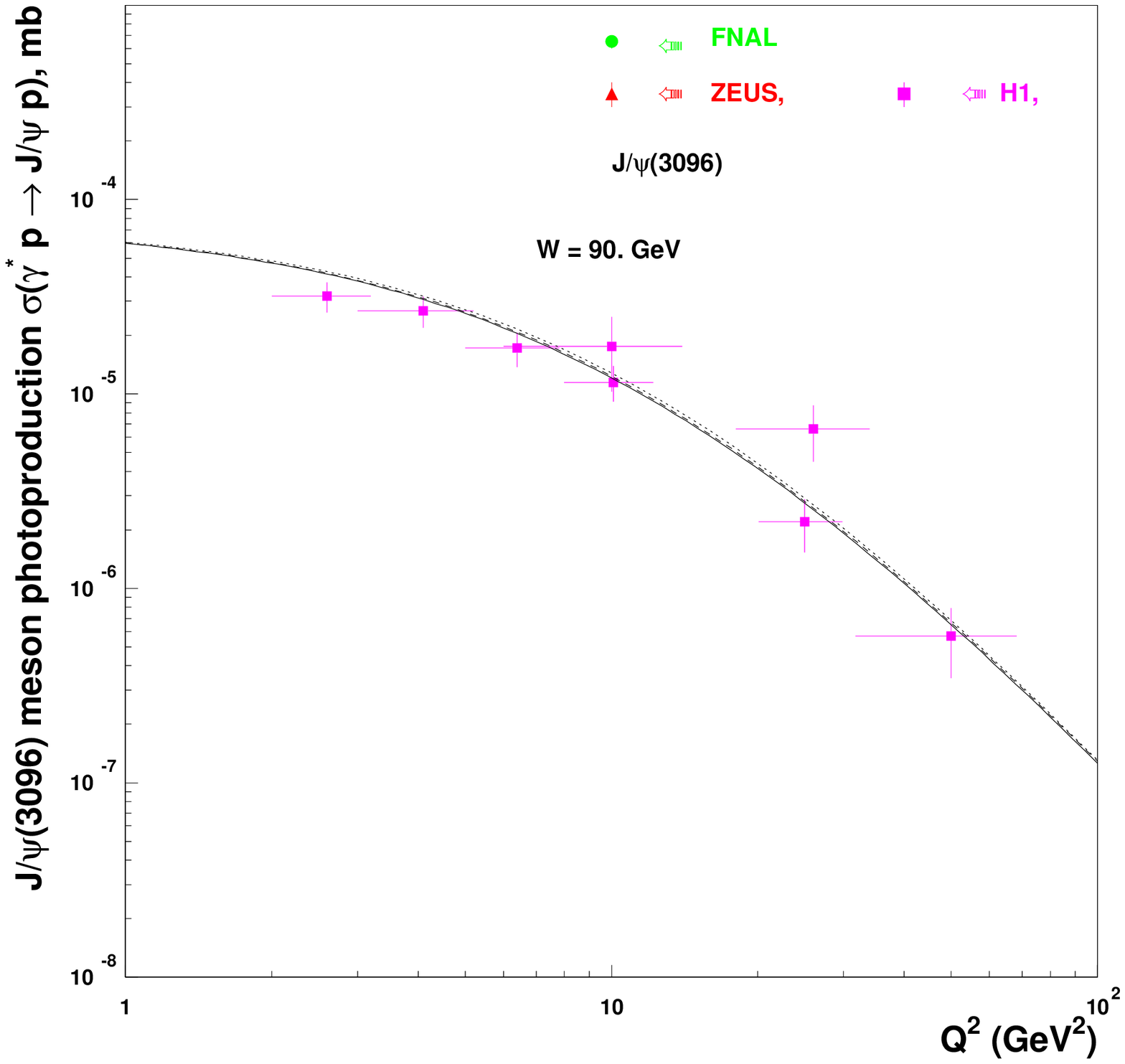}}

\vspace*{-1.3cm}
\parbox[t]{7.7cm}{\caption{Elastic cross section of exclusive $J/\psi$
virtual photoproduction as a function of $W$
for various $Q^2$. \label{fig:jpsi}}}
\hfill~\parbox[t]{7.7cm}{\caption{Elastic cross section of
exclusive $J/\psi$ virtual photoproduction as a
function of $Q^2$ for $W=90\; GeV$. \label{fig:jpsiq90}}}
\end{figure}



\begin{figure}[H]
\vspace*{-0.5cm}
\parbox[c]{8.7cm}{\epsfxsize=66mm
\epsffile{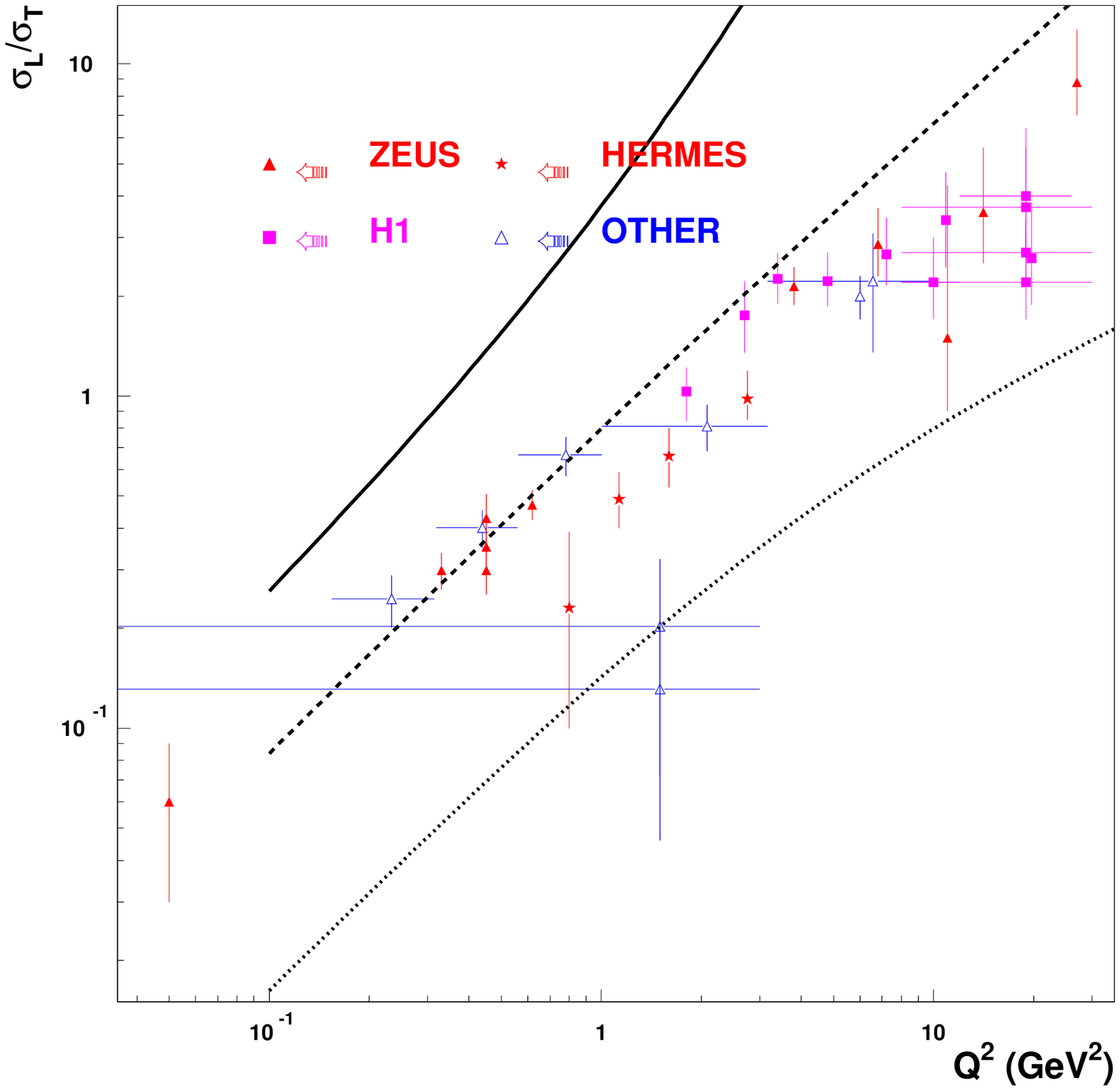}} \hfill~\parbox[c]{7.6cm}{\epsfxsize=66mm
\epsffile{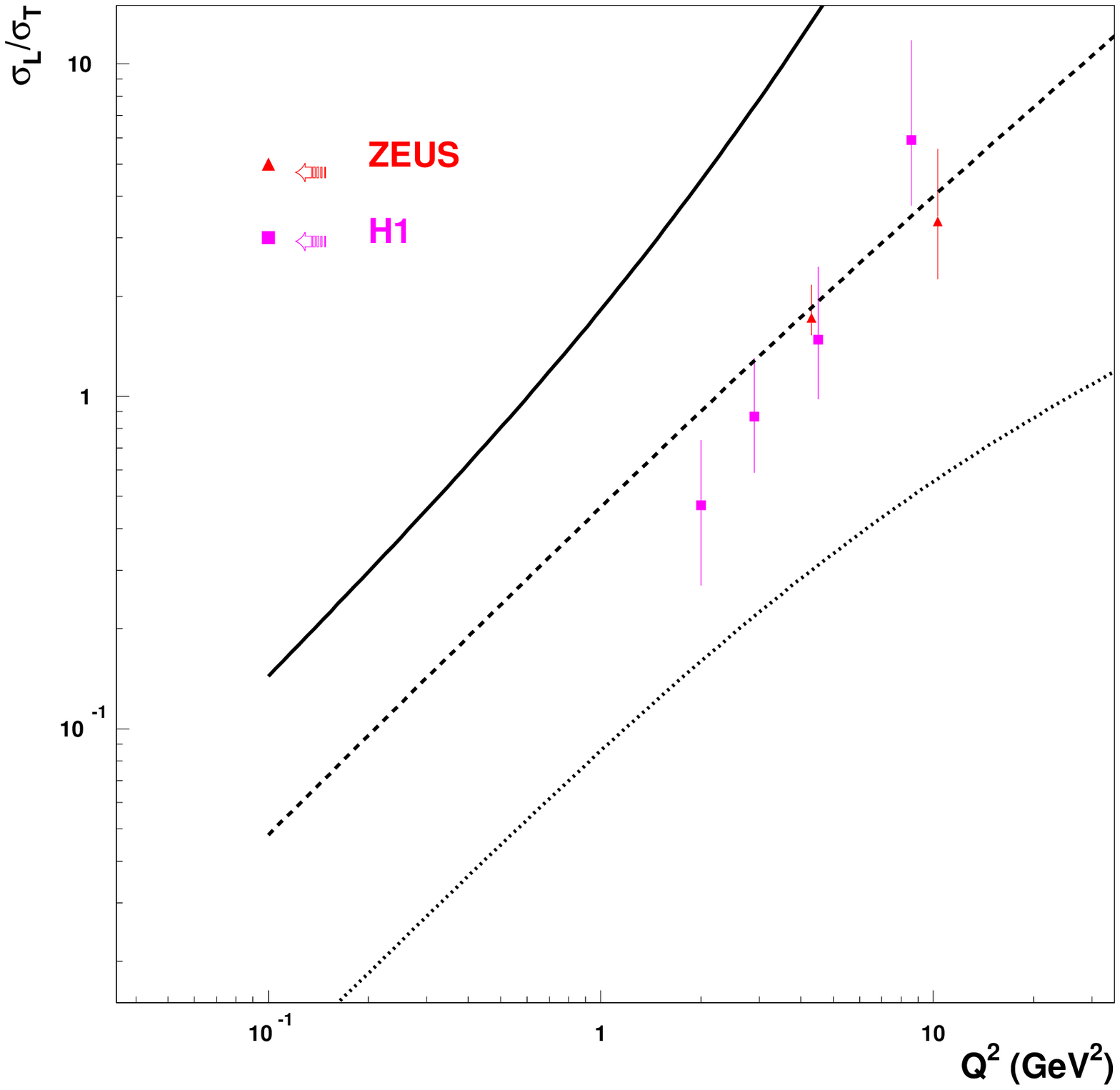}}

\vspace*{-1.7cm}
\parbox[t]{7.7cm}{\caption{Ratio of $\sigma_L/\sigma_T$ for exclusive
$\rho_0$ large $Q^2$ photoproduction.
\label{fig:rholt}}}
\hfill~\parbox[t]{7.7cm}{\caption{Ratio of $\sigma_L/\sigma_T$ for
exclusive $\varphi$ large $Q^2$ photoproduction.
\label{fig:philt}}}

\vspace*{-.7cm}

\parbox[c]{8.7cm}{\epsfxsize=66mm
\epsffile{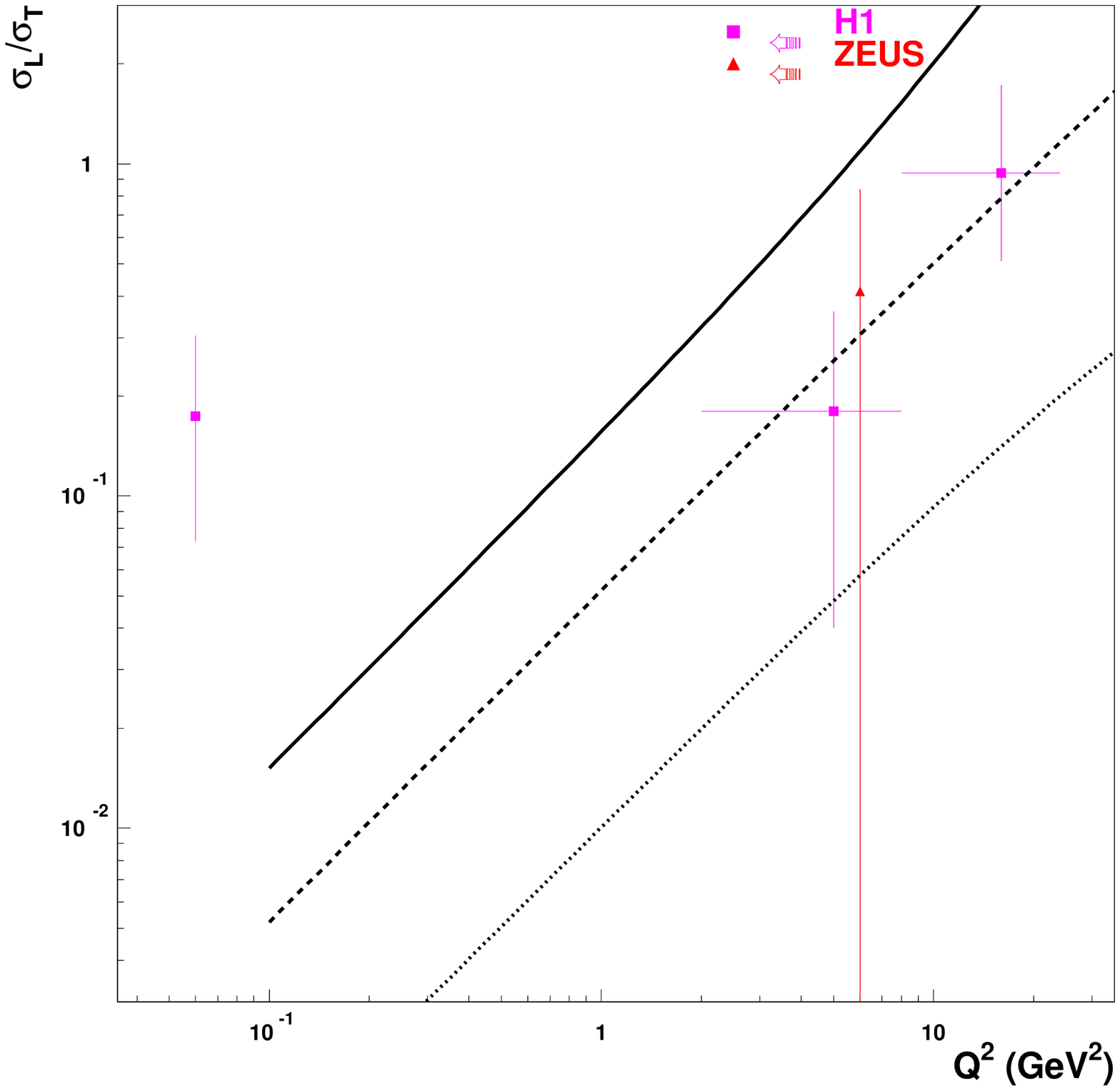}}

\vspace*{-1.7cm}
\parbox[t]{7.7cm}{\caption{Ratio of $\sigma_L/\sigma_T$ for exclusive
$J/\psi$ large $Q^2$ photoproduction.
\label{fig:jpsilt}}}

\end{figure}

We now plot the various ratios $\sigma_L/\sigma_T$ (these data were not fitted)
corresponding to
Eqs
(\ref{eq:couplingsChoice}),(\ref{eq:couplings1}),(\ref{eq:couplings2})
(shown with the solid ({\it choice I}), dashed ({\it choice II})
and dotted ({\it choice III}) lines) in Fig. \ref{fig:rholt},
\ref{fig:philt}, \ref{fig:jpsilt}.  The result shows, indeed, a
rapid increase of $\sigma_L/\sigma_T$ with increasing $Q^2$,
however one can see that our intermediate {\it choice II} is preferable to either {\it I} or {\it III} on this basis.

Let us examine the obtained dependences. We find that the data
prefer 
\be
R(Q^2\rightarrow \infty) \sim
\Big(\frac{Q^2}{M_V^2}\Big)^{n_1}\; , 
\ee where $n_1\simeq 2,\;
1,\; 0.3$ in {\it choice I, II and III}. Our, probably
oversimplified, estimates and the data show $0.3<n_1<1$, see
Fig. \ref{fig:rholt}, thus $\sigma\sim 1/Q^N$ where 
$N\in (4,4.4)$ as $N=6-2n_{1}$
for the {\it choice II} and  $N=5-2n_{1}$
for the {\it choice III}. However it is evident that 
new more precise data on $R$ are need.

\section{Conclusion}

We have shown that minor changes in the Soft Dipole Pomeron
model recently developed \cite{owrmodel} for vector meson
photoproduction allow us to describe well the new ZEUS data
\cite{NEWZEUS} on the differential and integrated cross-sections
for $\gamma p\to J/\psi p$. Again, all available data on
photoproduction of other vector mesons at $Q^{2}=0$ as well as
$Q^{2}\neq 0$ are well reproduced.

The changes made do not affect the main properties of the
model such as:
(i) the Pomeron intercept which is equal to one, (ii) the hardness of
the Pomeron,
{\it i.e.} the fact that it is a double pole in the complex $j$-plane.

We take directly into account the kinematical limits
through the variable  $z\propto
\cos\theta_{t}$. The nonlinear Pomeron trajectory
$\alpha_{P}(t)=1+\gamma\
(\sqrt{4m_{\pi}^{2}}-\sqrt{4m_{\pi}^{2}-t})$ turns out to be more
suitable for the nonlinearity of the diffractive cone
shown by the new ZEUS data. This is not unexpected as the linear behaviour
of the $\Pom$ trajectory is hard to reconcile with analyticity.
We have implemented also the correct limits
of $t$-integration. The last circumstance allows us to account for the
threshold behaviour of the cross-sections.

We would like to emphasize the following important points
(confirming the main findings of \cite{owrmodel} and repeating
some of them)
\begin{enumerate}
\item The new ZEUS data \cite{NEWZEUS} (in contrast to the old ones)
quite definitely point towards the nonlinearity of the Pomeron 
slope and trajectory.

\item Our model describes the data also at low energies due to
the kinematical shrinkage of the available $t$ region. This is
particularly important for $J/\psi$ production where the bulk of
the available data is not so far from its threshold.

\item Phenomenologically we find that in the region of available
$Q^2$ the ratio $ \sigma_L/\sigma_T  \sim
({Q^2}/{M_V^2})^{n_1}\; , $ where $0.3<n_1< 1\;$. The
definite conclusion can be derived only with new precise
data on the ratio $\sigma_L/\sigma_T$, especially for high
$Q^{2}$.

\item Pomeron and secondary Reggeons appear as universal objects
in Regge theory \cite{ref:PredazziBarone}. The corresponding 
$j$-singularities of the
$\gamma$ p amplitudes and their trajectories are universal. They
do not depend on the properties of the external particles and,
consequently, on $Q^{2}$ (only residues or vertex functions may
depend on $Q^{2}$). We believe that the unitarity restrictions
on the Pomeron contribution obtained strictly for the $hh$ case 
must hold also for $\gamma h$ if it is universal.

\item The growth with energy of hadronic total cross sections
and the restriction on the Pomeron intercept
($\alpha_{\Pom}(0)\leq 1$) implied by the Froissart-Martin
bound \cite{ref:MartinF}
imply that the Pomeron is a more complicated singularity than a
simple pole with $\alpha_{\Pom}(0)=1$. We have considered the
simplest case when the Pomeron is a double $j$-pole leading to
$\sigma(s)\propto \ln s$. We have shown that one does not need a
contribution with $\alpha(0)>1$ (hard Pomeron) violating
unitarity in order to describe the exclusive photoproduction
data in the present region of $Q^{2}$ and $t$.

\end{enumerate}

\subsection*{Acknowledgement}
We would like to thank Michele Arneodo, Alexander Borissov,
Jean-Rene Cudell and
Alessia Bruni for various and fruitful
discussions. One of us (E.M.) would like to thank the 
Department of Theoretical Physics of the University of Torino
for its hospitality and financial support during his visit
to Turin.


\end{document}